\documentclass[runningheads, envcountsame, a4paper]{llncs}
\usepackage{amssymb}
\usepackage{mathtools}
\usepackage{boxedminipage}
\usepackage{url}

\begin{document}
\newcommand{\bs}{\mathcal{B}} 
\newcommand{\np}{\mathbf{0}} 
\newcommand{\pc}{\ |\ } 
\newcommand{\lpc}{\ \Big|\ } 
\newcommand{\ip}[2]{#1( #2 )} 
\newcommand{\op}[2]{\overline{#1}\langle #2 \rangle} 
\newcommand{\ol}[1]{\overline{#1}} 
\newcommand{\subs}[2]{\{#1/#2\}} 

\newcommand{\nt}{N} 
\newcommand{\at}{M} 
\newcommand{\pbt}[1]{[\![#1]\!]} 

\newcommand{\dotsim}{\ \dot{ \sim }\ } 
\newcommand{\ra}{\rightarrow} 
\newcommand{\Ra}{\Rightarrow} 
\newcommand{\xra}[1]{\xrightarrow{#1}} 
\newcommand{\xRa}[1]{\xRightarrow{#1}} 

\newcommand{\lc}{::} 
\newcommand{\dom}[1]{\texttt{dom}(#1)} 
\newcommand{\len}[1]{\texttt{|}#1\texttt{|}} 
\newcommand{\kw}[1]{\texttt{#1}} 
\newcommand{\mt}{\mapsto} 
\newcommand{\de}{\stackrel{def}{=}} 
\newcommand{\cf}[1]{\langle #1 \rangle} 
\newcommand{\tts}[1]{\texttt{\textsc{#1}}} 
\newcommand{\laop}{\texttt{<\!-}} 
\newcommand{\raop}{\texttt{-\!>}} 

\newcommand{\vs}{\vspace{1mm}} 

\newcommand{\nl}{\kw{nil}} 
\newcommand{\tsg}[1]{[\![#1]\!]_g} 
\newcommand{\tsgs}[1]{\mathcal{T}_{#1}} 
\newcommand{\lag}{\hookrightarrow_g} 
\newcommand{\lagx}[1]{\xhookrightarrow{#1}_g} 
\newcommand{\letg}{\mapsto_g} 
\newcommand{\letgx}[1]{\xmapsto{#1}_g} 
\newcommand{\gag}{\rightarrow_g} 
\newcommand{\gagx}[1]{\xrightarrow{#1}_g} 
\newcommand{\gagX}[1]{\xRightarrow{#1}_g} 
\newcommand{\kt}[1]{\underline{#1}} 

\newcommand{\gotype}[1]{\mathfrak{{#1}}} 
\newcommand{\gl}[1]{\mathtt{#1}} 

\newcommand{\lae}{\hookrightarrow_e} 
\newcommand{\laex}[1]{\xhookrightarrow{#1}_e} 
\newcommand{\gae}{\rightarrow_{e}}
\newcommand{\gaex}[1]{\xrightarrow{#1}_{e}}
\newcommand{\gaeX}[1]{\xRightarrow{#1}_e} 
\newcommand{\tse}[1]{[\![#1]\!]_e} 
\newcommand{\tses}[1]{\mathcal{T}_{#1}} 
\newcommand{\mb}{\mathfrak{m}} 
\newcommand{\MB}{\mathcal{M}} 
\newcommand{\erl}[1]{\mathtt{#1}} 


\title{The Buffered $\pi$-Calculus: A Model for Concurrent Languages}
\titlerunning{The Buffered $\pi$-Calculus}

\author{Xiaojie Deng\inst{1} \and Yu Zhang\inst{2} \and Yuxin Deng\inst{1} \and Farong Zhong\inst{3}}
\authorrunning{X. Deng, Y. Zhang, Y. Deng, and F. Zhong}

\institute{Department of Computer Science and Engineering, Shanghai Jiao Tong University, Shanghai, China
  \email{$\{$lvchaxj, yuxindeng$\}$@gmail.com}
\and Institute of Software, Chinese Academy of Sciences, Beijing, China
  \email{yzhang@ios.ac.cn}
\and Department of Computer Science, Zhejiang Normal University, Zhejiang, China
  \email{zfr@zjnu.cn}
}

\maketitle
\setcounter{footnote}{0}


\begin{abstract}
Message-passing based concurrent languages are widely used in developing large distributed and coordination systems.
This paper presents the buffered $\pi$-calculus --- a variant of the $\pi$-calculus where channel names
are classified into buffered and unbuffered: communication along buffered channels is asynchronous,
and remains synchronous along unbuffered channels.
We show that the buffered $\pi$-calculus can be fully simulated in the polyadic $\pi$-calculus with respect to strong bisimulation.
In contrast to the $\pi$-calculus which is hard to use in practice, the new language enables easy and clear modeling
of practical concurrent languages. We encode two real-world concurrent languages in the buffered $\pi$-calculus:
the (core) Go language and the (Core) Erlang. Both encodings are fully abstract with respect to weak bisimulations.

\end{abstract}


\section{Introduction}
\label{sec:intro}

Concurrent programming languages become popular in recent years thanks to the large demand of distributed computing and
the pervasive exploitation of multi-processor architectures. Unlike the shared-memory concurrency model, which is now mainly
used on multi-processor platforms, message passing based concurrent languages are particularly popular in developing
large distributed, coordination systems.
Indeed, quite a few real-world concurrent languages are intensively used in industry.
The most well-known languages are probably Erlang, developed by Ericsson~\cite{Armstrong1997},
and the much younger language Go, developed by Google~\cite{GoSpec2012}.
Both languages achieve their asynchronous communication via order-preserving message passing.

On the other side, the $\pi$-calculus~\cite{Milner1992,Sangiorgi2001} has shown its success
in modeling and verifying both specifications and implementations.
Its asynchronous variant~\cite{Boudol1992,Honda1991} is a good candidate as the target formal model.
Despite the fact that it is called asynchronous, communication in the asynchronous $\pi$-calculus is however synchronous.
It is shown in \cite{Beauxis2008} that the communication modelled by the asynchronous $\pi$-calculus
is equivalent to message passing via bags --- senders put messages into some bags, and
receivers may get arbitrary messages from these bags.
This result indicates that additional effort should be made to respect the order of the messages,
which is adopted in the implementation of many concurrent languages.

In view of this, we may expect a formal model where asynchronous communication is supported natively.
In fact, our primary goal is to achieve a formal model by which we can easily define a formal semantics of Go
and do verification on top of it.
The developers of Go claim that the concurrency feature of Go is rooted in CSP~\cite{Hoare1978},
while we show that the $\pi$-calculus should be an appropriate model for Go
as CSP does not support transmission of channel names over channels.

In the spirit of the name passing mechanism of the $\pi$-calculus and the channel type of the Go language,
we extend the $\pi$-calculus by introducing a special kind of names, each associated with a first-in-first-out buffer.
We call these names {\em buffered names}. Communication along buffered names is asynchronous,
while that along unbuffered (normal) names remains synchronous.
We call this variant language the buffered $\pi$-calculus, and abbreviate it as the $\pi_b$-calculus.

We develop the $\pi_b$-calculus by defining its operational semantics as a labelled transition system and
supplying an encoding into the polyadic $\pi$-calculus.
We also present translations of the languages Go and Erlang into the $\pi_b$-calculus
and show that the model is sufficient and relatively easier for modeling real-world concurrent languages.

\subsection{Related Work}

Beauxis \emph{et al} introduced the $\pi_{\mathfrak{B}}$-calculus in order to
study the asynchronous nature of the asynchronous $\pi$-calculus~\cite{Beauxis2008}.
Their asynchronous communication is achieved via explicit use of buffers.
In case that the buffers are ordered structures such as queues or stacks, the asynchronous communication
modelled by $\pi_{\mathfrak{B}}$ differs from that by the asynchronous $\pi$-calculus.
While communication in the $\pi_{\mathfrak{B}}$-calculus is always asynchronous,
we keep both synchronous and asynchronous communication in the $\pi_b$-calculus, through different types of names.

Encoding programming languages in process calculus have been studied by many researchers.
Milner defines the semantics of a non-trivial parallel programming language by a translation into CCS in \cite{Milner1989}.
In \cite{Walker1995}, a translation from a parallel object oriented language to the minimal $\pi$-calculus is presented.
The correctness of the translation is justified by the operational correspondence between units and their encodings.
Our treatments to the Go language follows the approach in \cite{Walker1995}.
In addition, we show a full abstraction theorem, namely equivalent Go programs are translated into equivalent $\pi_b$ processes.

For functional languages, Noll and Roy~\cite{Noll2005} presented an initial translation mapping
from a Core Erlang~\cite{Carlsson2000} to the asynchronous $\pi$-calculus.
Later on they~\cite{Roy2006} improved the translation by revising the non-deterministic encoding of pattern matching based expressions,
and by adding the encoding for tuples.
Their translations, however, are not sound in the sense that the order of messages is not always respected.
By modelling the mailbox structure explicitly by buffered names in the $\pi_b$-calculus,
we obtain a more accurate encoding which is fully abstract with respect to weak bisimulation.

\subsection{Outline}
The rest of the paper is structured as follows.
Section~\ref{sec:pib} presents the syntax and semantics of the $\pi_b$-calculus and a simple encoding
in the polyadic $\pi$-calculus~\cite{Milner1991}. We show that this encoding preserves the strong bisimulation relation.
In Section~\ref{sec:go} we define a formal semantics for Go and present an encoding of Go in the $\pi_b$-calculus.
Section~\ref{sec:erlang} is devoted to Core Erlang, in which an improved encoding is presented.
And finally, Section~\ref{sec:con} concludes the paper.


\section{The $\Pi_b$-Calculus}
\label{sec:pib}

We assume an infinite set $\mathcal{N}$ of names, ranged over by $a, b, c, d, x, y$.
%
Processes are defined by the following grammar.
\[ P, Q, \ldots := \sum_{i \in I}\pi_i.P_i \lpc P \pc Q \lpc (\nu c:n)P \lpc (\nu c)P \lpc !P \]
where $\pi = \ip{c}{x} \pc \op{c}{d} \pc \tau$.

Most of the syntax is standard:
$\sum_{i \in I}\pi_i.P_i$ is the guarded choice ($I$ is finite), which behaves nondeterministically
as one of its components $\pi_j.P_j$ for some $j \in I$; composition $P|Q$ acts as $P$ and $Q$ running in parallel;
$!P$ is the replication of process $P$;
Prefixes $\ip{c}{x}$ and $\op{c}{d}$ are input and output along name $c$; and $\tau$ is the silent action.
We write $\np$ for the empty guarded choice, it is the process which can do nothing.

The $\pi_b$-calculus extends the $\pi$-calculus in the fact that names can be buffered or unbuffered.
Unbuffered names are names in the $\pi$-calculus, and buffered names have the buffer attribute specified by a \emph{buffer store}.
A buffer store, denoted by $\bs$, is a partial function from buffered names to pairs $(n, l)$,
where $n$ is a positive integer representing the capacity of the buffer,
and $l$ is a list of names in the buffer, with the same order.
Both $(\nu c) P$ and $(\nu c: n) P$ are called {\em new processes}.
The (standard) new process $(\nu c)P$ specifies that $c$ (whether buffered or unbuffered) is a local name in $P$.
The extended new process $(\nu c:n)P$ creates a local buffered name $c$,
whose associated buffer has the capacity $n$ for asynchronous communication inside $P$.
Notice that $(\nu c) P$ only says that the name $c$ is local and does not imply that $c$ is unbuffered ---
$c$ can be a buffered name whose buffer is already created in the buffer store.

Input process $\ip{c}{x}.P$ and output process $\op{c}{d}.P$ can communicate with each other along name $c$ when they run in parallel.
If $c$ is an unbuffered name, the communication is synchronous and happens as in the $\pi$-calculus: the object $d$ is passed from the
output side to the input side.
If $c$ is a buffered name, then the communication becomes asynchronous: the output process simply puts $d$ into the buffer of $c$
if it is not full and continues, or blocks if the buffer is full; the input process retrieves the oldest value from the buffer of $c$
if it is not empty and continues, or blocks if the buffer is empty.

As usual, we write $\tilde{c}$ for a sequence of names, and abbreviate $(\nu c_1) \dots (\nu c_n)P$ to $(\nu c_1 \dots c_n)P$.
A name $x$ is bound if it appears in input prefix, otherwise it is free.
We write $P\subs{\tilde{c}}{\tilde{x}}$ for the process resulting from simultaneously substituting $c_i$ for each free $x_i$ in $P$.
The newly created name $c$ in $(\nu c:n)P$ or $(\nu c)P$ are local names. A name is global if it is not localized
by any new operator. We use $ln(P)$ and $gn(P)$ for the set of local names and global names occurring in $P$.

Throughout the development of the paper, we assume the following {\em De Barendregt name convention}:
\begin{center}{\em Local names are different from each other and from global names.}\end{center}
For instance, we shall never consider processes like  $\op{a}{c}.(\nu a)P$ or $(\nu a)(\nu a)P$.
We note that this convention is dispensable and we simply adopt it to make the presentation of the calculus simple and clean.
One can also remove the convention and use syntactic rules to manage name conflicts, but dealing with names in buffers
can be very subtle.

A process can send a local name into a buffer.
The fact that a name stored in buffers is local must be tracked, because it may affect the name scope when another process
retrieves this name from the buffer. The convention also works for buffer stores.
We shall discuss more on this when defining the operational semantics.
Inside a buffer store, a value of the form $(\nu c)$ indicates that the name $c$ was sent into the buffer when it was local.
Given a buffer store $\bs$, we write $gn(\bs(b))$ for the set of global names that occur in $b$'s buffer,
and $gn(\bs) = \bigcup_{b \in \dom{\bs}}gn(\bs(b))$. Similarly $ln(\bs(b))$ and $ln(\bs)$ for local names in $\bs(b)$ and $\bs$.
The buffer store $\bs\subs{c}{d}$ is obtained by substituting $c$ for each $d$ in $\bs$.

We say a process $Q$ is guarded in $P$, if every occurrence of $Q$ in $P$ is within some prefix process.
Intuitively, a guarded process cannot affect the behavior of its host process until the action induced by its guarding prefix is performed.
New operators are guarded in $P$ if all new processes are guarded in $P$.

The structural congruence $\equiv_\bs$ with respect to the buffer store $\bs$ is defined as the smallest congruence relation
over processes satisfying the laws in Table~\ref{tab:pbsc}.
\begin{table}
\begin{enumerate}  \setlength{\itemsep}{0ex}
  \item $P \equiv_\bs Q$, if $Q$ is obtained from $P$ by renaming bound names, or local names not occurring in $\bs$.
  \item $P \pc Q \equiv_\bs Q \pc P;    P \pc (Q \pc R) \equiv_\bs (P \pc Q) \pc R;    P \pc \np \equiv_\bs P $.
  \item $!P \equiv_\bs P \pc !P$.
  \item $(\nu c)(\nu d)P \equiv_\bs (\nu d)(\nu c)P$.
  \item $(\nu c)\np \equiv_\bs \np, \text{ if } c \not\in ln(\bs)$.
  \item $(\nu c)(P \pc Q) \equiv_\bs (\nu c)P \pc Q, \text{ if } c \not\in ln(\bs) \wedge c \not\in gn(Q)$
\end{enumerate}
\caption{Structural Congruence}
\label{tab:pbsc}
\end{table}
Structural congruence allows us to pull unguarded new operators to the ``outermost'' level.

Buffer store $\bs$ is \emph{valid} for process $P$ if each local name of $\bs$ appears in some new operator occurring
at the outermost level of $P$, i.e., for every $c \in ln(\bs)$, $P \equiv_\bs (\nu c)P'$ for some $P'$.

\subsection{Operational Semantics}

\begin{table}
  \[ \begin{array}{l}
      \tts{IU}\ \cfrac{c \not \in \dom{\bs}}{\ip{c}{x}.P, \bs \xra{\ip{c}{d}} P\subs{d}{x}, \bs} \quad
      \tts{OU}\ \cfrac{c \not \in \dom{\bs}}{\op{c}{d}.P, \bs \xra{\op{c}{d}} P, \bs} \quad
      \tts{Open}\ \cfrac{ P, \bs\subs{c}{\nu c} \xra{\op{d}{c}} P', \bs' }{ (\nu c)P, \bs \xra{\op{d}{\nu c}} P', \bs'}
      \vs \\

      \tts{IB}\ \cfrac{ \bs(b) = (n, [d] \lc l) }{ \ip{b}{x}.P, \bs \xra{\tau} P\subs{d}{x}, \bs[b \mt (n, l)] } \qquad\ \,
      \tts{OB}\ \cfrac{ \bs(b) = (n, l);\ \len{l} < n }{ \op{b}{d}.P, \bs \xra{\tau} P, \bs[b \mt (n, l \lc [d])] }
      \vs \\

      \tts{IBG}\ \ \cfrac{ \bs(b) = (n, l);\ \len{l} < n;\ b \not\in ln(P) }{ P, \bs \xra{\ip{b}{d}} P, \bs[b \mt (n, l \lc [d])] } \qquad
      \tts{OBG}\ \ \cfrac{ \bs(b) = (n, [d] \lc l);\ b \not\in ln(P) }{ P, \bs \xra{\op{b}{d}} P, \bs[b \mt (n, l)] }
      \vs \\

      \tts{Sum}\ \ \cfrac{ j \in I;\ \pi_j.P_j, \bs \xra{\alpha} P', \bs' }{ \sum_{i \in I}\pi_i.P_i, \bs \xra{\alpha} P', \bs'} \qquad
      \tts{Com}\ \ \cfrac{ P, \bs \xra{c(d)} P', \bs;\ Q, \bs \xra{\op{c}{d}} Q', \bs;\ c \not\in \dom{\bs}}{ P \pc Q, \bs \xra{\tau} P'\pc Q', \bs}
      \vs \\

      \tts{Par}\ \ \cfrac{ P, \bs \xra{\alpha} P', \bs';\ \text{new operators are guarded in $P \pc Q$}}{ P \pc Q, \bs \xra{\alpha} P' \pc Q, \bs'}
      \vs \\

      \tts{New}\ \ \cfrac{ P, \bs\subs{c}{\nu c} \xra{\alpha} P', \bs';\ c \not\in n(\alpha) }{ (\nu c)P,\bs \xra{\alpha} (\nu c)P',\bs'\subs{\nu c}{c} } \qquad
      \tts{Stru}\ \ \cfrac{ P \equiv_\bs P';\ P', \bs \xra{\alpha} Q', \bs';\ Q' \equiv_{\bs'} Q }{ P, \bs \xra{\alpha} Q, \bs' }
      \vs \\

      \tts{NewB}\ (\nu b:n)P,\bs \xra{\tau} (\nu b)P,\bs[b \mt (n,[\ ])]
  \end{array} \]
  \caption{Operational Semantics of $\pi_b$}
  \label{tab:pbos}
\end{table}

The (early) transition semantics of $\pi_b$ is given in terms of a labelled transition system generated
by the rules in Table~\ref{tab:pbos}.
The transition rules are of the form $P, \bs \xra{\alpha} P', \bs'$, where $P, P'$ are processes, $\bs, \bs'$ are buffer stores
and $\alpha$ is an action,  which can be one of the forms:
the silent action $\tau$, free input $\ip{c}{d}$, free output $\op{c}{d}$ or bound output $\op{c}{\nu d}$.
We write $n(\alpha)$ for the set of names occurring in $\alpha$.

These rules are compatible with the transition rules for the $\pi$-calculus.
\tts{IU} and \tts{OU} are rules for unbuffered names and synchronous communication is specified by \tts{Com}.
\tts{IB} and \tts{OB} define the asynchronous communication along buffered names:
$\ip{b}{x}.P$ performs a $\tau$ action by receiving the ``oldest'' name $d$ from $b$'s buffer,
while $\op{b}{d}.P$ performs a $\tau$ action by inserting $d$ into $b$'s buffer.
Communication along buffered names is asynchronous because it involves two transitions (\tts{IB} and \tts{OB})
and other actions may occur between them.

\tts{IBG} and \tts{OBG} indicate that a buffer store itself may have actions.
If $b$ is a global buffered name, that is $(\nu b)$ does not occur in $P$, then we can insert names to or receive names
from $b$'s buffer directly.
In \tts{New} and \tts{Open}, the substitutions on the buffer store are for the sake of validity.
\tts{NewB} is the rule for the extended new process. After creating an empty buffer for $b$,
the capacity parameter $n$ is dropped, leaving the new operator indicating that $b$ is a local name.

The \tts{Par} rule describes how processes can progress asynchronously, which typically happens with buffered names.
However, unlike in the $\pi$-calculus, where we have open/close rules to manage name scope extension,
in the $\pi_b$-calculus, it is hard (perhaps impossible) to define an appropriate close rule because when a local name
is exported to a buffer, it becomes hard to track which process will retrieve the name so as to determine the name scope.
For instance, consider the process $P_1 | P_2 | P_3$ where $P_1 = (\nu a)\op{b}{a}.P'_1$, $P_2 = \ip{b}{y}.\cdots$,
$P_3 = \ip{b}{z}.\cdots$ and a valid buffer store $\bs = [b \mt (2, [\ ])]$.
In the $\pi_b$-calculus, $P_1$ inserts the local $a$ into $b$'s buffer by a $\tau$ action,
then it can possibly be received by $P_2$ or $P_3$, hence tracking the scope of $a$ becomes very hard.
Our solution here is to prevent processes from inserting local names into buffers when they are running in parallel with other processes.
For processes like the above example, we extend the scope of $a$ to the entire process by structural congruence laws
and obtain a process in the form $ (\nu a)(\op{b}{a}.P'_1 | P_2 | P_3)$ thanks to the name convention.
This avoids the scope problem.

We have adopted the name convention which simplifies the definition of the labeled transition system.
Dealing with names with buffers is subtle and the transition rules without the name convention are discussed in the next subsection.

The following proposition says that transition rules preserve buffer validity:
\begin{proposition}
If $\bs$ is valid for process $P$ and we have the transition $P, \bs \xra{\alpha} P', \bs'$, then $\bs'$ is valid for $P'$.
\end{proposition}

As in the $\pi$-calculus, strong bisimulation over the set of $\pi_b$ processes can be defined as follows.
\begin{definition}
  A symmetric binary relation $\mathcal{R}$ over $\pi_b$ processes is a \emph{bisimulation}, if whenever $(P, \bs_P) \mathcal{R} (Q, \bs_Q)$ and $(P, \bs_P) \xra{\alpha} (P', \bs'_P)$,
  \[\exists (Q', \bs'_Q) \ . \ (Q, \bs_Q) \xra{\alpha} (Q', \bs'_Q) \wedge (P', \bs'_P) \mathcal{R} (Q', \bs'_Q) \]
  Strong bisimilarity $\dot{\sim}$ is the largest strong bisimulation over the set of $\pi_b$ processes. $(P, \bs_P)$ and $(Q, \bs_Q)$ are strongly bisimilar, written as $(P, \bs_P)\ \dot{\sim}\ (Q, \bs_Q)$, if they are related by some strong bisimulation.
\end{definition}

\subsubsection{Transition Rules without Name Conversion}
As mentioned above, some transition rules require extra conditions to deal with name conflict without the name conversion.
These rules are shown in Table~\ref{tab:pbos2}.

\begin{table}
  \[ \begin{array}{l}
    \tts{New*}\ \ \cfrac{ P, \bs\subs{c}{\nu c} \xra{\alpha} P', \bs';\ c \not\in n(\alpha);\ c \not\in gn(\bs);\ c \not\in ln(\bs') }
    { (\nu c)P,\bs \xra{\alpha} (\nu c)P',\bs'\subs{\nu c}{c} }
    \vs \\

    \tts{NewB*}\ \cfrac{ b \not\in \dom{\bs} }{ (\nu b:n)P,\bs \xra{\tau} (\nu b)P,\bs[b \mt (n,[\ ])]}
    \vs \\

    \tts{Open*}\ \cfrac{ P, \bs\subs{c}{\nu c} \xra{\op{d}{c}} P', \bs';\ c \not\in gn(\bs) }
    { (\nu c)P, \bs \xra{\op{d}{\nu c}} P', \bs'}
  \end{array} \]
  \caption{Operational Semantics without Name Conversion}
  \label{tab:pbos2}
\end{table}

The problem is how can we determine a local name in the buffers refers to which local name of the process.
For instance, suppose $P = (\nu c)(\nu c)P'$ with valid buffer store $\bs = [b \mt (5, [\nu c])]$.
We have no idea the $\nu c$ in $\bs$ refers to which one of the two local $c$s in $P$.
Therefore we first assume the local names in buffers are localized by the ``outermost'' and ``\emph{leftmost}'' new operator of the process,
and add additional conditions to transition rules to respect the assumption.

In \tts{New*}, for those global $c$ in $\bs$, they are semantically different from the local $c$ in $P$.
For example $P = \op{b}{c}.(\nu c)\op{b}{c}$, with $\bs = [b \mt (5, [\ ])]$.
A $\tau$ transition inserts the global $c$ into $b$'s buffer, and we have $P' = (\nu c)\op{b}{c}$, with $\bs' = [b \mt (5, [c])]$.
At this point, we intend to insert the local $c$ into $b$'s buffer, this local $c$
is apparently different from the global $c$ already in the buffer.
We add the condition $c \not\in gn(\bs)$ to enforce an renaming of the local $c$ of $P$ before the insertion.
The same discussion applies to the same extra condition of \tts{Open*}.

In rare cases, another condition is required for \tts{New*}.
Suppose, for instance, $P = (\nu c)(\nu c)\op{b}{c}.\op{b}{c}$, with $\bs = [b \mt (5, [\ ])]$.
According to existing \tts{New} rule,
the process may perform a $\tau$ action inserting a local $c$ into $b$'s buffer and become $(\nu c)(\nu c)\op{b}{c}$,
with buffer store changed to $\bs[b \mt (5, [\nu c])]$ where the local $c$ is actually localized by the second $(\nu c)$.
But by our assumption, it would refer to the first local $c$ of the process.
We avoid this inconsistency by introducing the condition $c \not\in ln(\bs')$.

The condition $b \not\in \dom{\bs}$ in \tts{NewB*} guarantees that
a fresh buffered name is used.

\subsubsection{Examples}

We demonstrate these transition rules by showing some examples.
The following example illustrates the asynchronous communication by buffered names.

\begin{example}
$P = (\nu a)\op{b}{a}.a(x) \pc b(y).\op{y}{c} \pc R$ and $\bs = [b \mt (5, [\ ])]$
\[ \begin{array}{llll}
    & & (\nu a)\op{b}{a}.a(x) \pc b(y).\op{y}{c} \pc R,   & \bs[b \mt (5, [\ ])] \\
     &\equiv_{\bs} & (\nu a)(\op{b}{a}.a(x) \pc b(y).\op{y}{c} \pc R),   & \bs[b \mt (5, [\ ])] \\
    \tts{New} &\xra{\tau} & (\nu a)(a(x) \pc b(y).\op{y}{c} \pc R),   & \bs[b \mt (5, [\nu a])] \\
    \tts{New} &\xra{\tau} & (\nu a)(a(x) \pc \op{a}{c} \pc R),   & \bs[b \mt (5, [\ ])] \\
\end{array} \]
\end{example}

$P$ is a parallel composition, we may not use \tts{Par} immediately as it contains an unguarded new process.
After moving the new operator to the outermost level, we may apply the \tts{New} rule
which induces a $\tau$ transition `sending' the local $a$ into $b$'s buffer.
Notice that the local $a$ is not directly inserted into the buffer, but in a substituting way ---
$a$ is inserted into $b$'s buffer in the premise of \tts{New}, then $a$ is replaced by $\nu a$ in the conclusion.
The second $\tau$ step describes the `receiving' of a local name from a buffer.

The next three examples illustrate the extra \tts{New*} rule.
\begin{example}
$P = \op{b}{a}.(\nu a)\op{b}{a}.\op{b}{a}$ and $\bs = [b \mt (5, [\ ])]$
\[ \begin{array}{llll}
    && \op{b}{a}.(\nu a)\op{b}{a}.\op{b}{a},   & \bs[b \mt (5, [\ ])] \\
    \tts{IB}  & \xra{\tau} & (\nu a)\op{b}{a}.\op{b}{a},   & \bs[b \mt (5, [a])] = \bs' \\
       &\equiv_{\bs'} & (\nu a')\op{b}{a'}.\op{b}{a'},   & \bs[b \mt (5, [a])] \\
    \tts{New*}  &\xra{\tau} & (\nu a')\op{b}{a'},   & \bs[b \mt (5, [a, \nu a'])] \\
    \tts{New*}  &\xra{\tau} & (\nu a')\np,   & \bs[b \mt (5, [a, \nu a', \nu a'])] \\
\end{array} \]
\end{example}
The first $\tau$ step follows from \tts{IB}.
At this point, the local name $a$ occurs free in buffer store, hence an $\alpha$-conversion is required.
After renaming $a$ to $a'$, a second $\tau$ transition `sending' the local name to $b$'s buffer.

\begin{example}
$P = (\nu a)\op{b}{a}.(\nu a)\op{b}{a}$ and $\bs = [b \mt (5, [\ ])]$
\[ \begin{array}{llll}
    && (\nu a)\op{b}{a}.(\nu a)\op{b}{a},   & \bs[b \mt (5, [\ ])] \\
    \tts{New*}  &\xra{\tau} & (\nu a)(\nu a)\op{b}{a},   & \bs[b \mt (5, [\nu a])] = \bs' \\
    &\equiv_{\bs'} & (\nu a)(\nu a')\op{b}{a'},   & \bs[b \mt (5, [\nu a])] \\
    \tts{New*}  &\xra{\tau} & (\nu a)(\nu a')\np,   & \bs[b \mt (5, [\nu a, \nu a'])] \\
\end{array} \]
\end{example}
After inserting the local $a$ into the buffer, the process contains two outermost new operators
and they are syntactically the same (semantically different).
Following \tts{New*}, we first determine the action of $(\nu a)\op{b}{a}$ with $\bs[b \mt (5, [a])]$.
Since the local name $a$ occurs free in the buffer store, an $\alpha$-conversion is required.
After that, a $\tau$ transition results in $(\nu a')\np$ with buffer store $\bs[b \mt (5, [a, \nu a'])]$.
Finally, all the $a$ in buffer store are modified back to $\nu a$.

\begin{example}
$P = (\nu a)(\nu a)\op{b}{a}.\op{b}{a}$ and $\bs = [b \mt (5, [\ ])]$
\[ (\nu a)\op{b}{a}.\op{b}{a}, \bs[b \mt (5, [\ ])] \quad \xra{\tau} \quad (\nu a)\op{b}{a}, \bs[b \mt (5, [\nu a])] \]
\[ \begin{array}{llll}
    && (\nu a)(\nu a)\op{b}{a}.\op{b}{a},   & \bs[b \mt (5, [\ ])] = \bs' \\
    &\equiv_{\bs'}& (\nu a)(\nu a')\op{b}{a'}.\op{b}{a'},   & \bs[b \mt (5, [\ ])] \\
    \tts{New*}  &\xra{\tau} & (\nu a)(\nu a')\op{b}{a'},   & \bs[b \mt (5, [\nu a'])] \\
    \tts{New*}  &\xra{\tau} & (\nu a)(\nu a')\np,   & \bs[b \mt (5, [\nu a', \nu a'])] \\
\end{array} \]
\end{example}
$P$ contains two outermost new operators at the beginning.
We first determine the action of $(\nu a)\op{b}{a}.\op{b}{a}$ with buffer $\bs[b \mt (5, [\ ])]$.
A $\tau$ transition leads to $(\nu a)\op{b}{a}$ with $\bs[b \mt (5, [\nu a])]$
where the local $a$ actually refers to the second $\nu a$ of the original process.
Hence an $\alpha$-conversion of the second $\nu a$ is required to distinguish itself with the first one.
This requirement is captured by $a \not\in ln(\bs')$ in \tts{New*}.

\subsubsection{An interlude of Close Rule and Structure Congruence}
In $\pi$ calculus, including which law in structure congruence and which other rule in transition rules is a trade-off. This phenomenon also exists in our $\pi_b$ calculus. One may ask for including a similar $\tts{Close}_\pi$ rule, which generates a $\tau$ action by synchronizing an input and a bound output action, and omitting those scope extension laws in structure congruence.
\[ \tts{Close}_{\pi} \quad \cfrac{ P \xra{\op{b}{\nu a}} P';\ Q \xra{\ip{b}{a}} Q'}{P \pc Q \xra{\tau} (\nu a)(P' \pc Q') } \]
In $\tts{Close}_{\pi}$, the local name moves to the action in the premise and to the outermost level in the conclusion.
What should the rule be in $\pi_b$?

Suppose $P$ performs a $\tau$ action sending a local name $a$ to $b$'s buffer and becomes $P'$,
and at some point in the future $Q$ performs a $\tau$ action receiving this local name from the buffer and becomes $Q'$,
the scope of $a$ should contain both $P'$ and $Q'$.
But if $P'$ derives to other process during the period between the two silent actions,
how can we determine which processes we should encompass by the new operator $\nu a$.

In the $\pi$-calculus, communications are synchronous, input process would proceed unless a
complement (bound) output process is ready, and vice verse.
However, in the $\pi_b$-calculus, communications along buffered names are asynchronous,
other actions may occur between the two transitions.
For this reason, we choose scope extension laws instead of some $\tts{Close}$ rules.

\subsection{Encoding in the Polyadic $\Pi$-Calculus}

We demonstrate an encoding of the $\pi_b$-calculus in the polyadic $\pi$-calculus.

Intuitively, a $\pi_b$ name $c$ is encoded into a pair of $\pi$ names $(c_1, c_2)$ by the injective \emph{name translation function} $\nt$.
In the name pair, $c_1$ is called the \emph{input name} and $c_2$ the \emph{output name} of $c$.
In addition, input and output names for unbuffered names are identical, but not for buffered names.
The two translation names of buffered name $b$ are exactly the names along which a buffer process
modelling the buffer of $b$ receives and sends values.
\begin{table}
    \begin{enumerate}
      \item If $a$ is a unbuffered name, then $a_1 = a_2$ where $\nt(a) = (a_1, a_2)$.
      \item If $b$ is a buffered name, then $b_1 \not= b_2$ where $\nt(b) = (b_1, b_2)$.
      \item For any two names $c \not= d$, $c_1 \not= d_1 \wedge c_1 \not= d_2 \wedge c_2 \not= d_1 \wedge c_2 \not= d_2$.
    \end{enumerate}
    \caption{Name Translation Function $\nt$}
    \label{tab:pbnt}
\end{table}

The buffer process is defined in Table~\ref{tab:pbbp}.
Intuitively speaking, $F_{n, L}(b_1, b_2)$ is the $\pi$ representation of $b$'s buffer, where $n$ denotes the capacity and $L$ is a list of $\pi$ name \emph{pairs}.
This process may further receive a pair of names along its second parameter $b_2$ if $L$ is not full ($\len{L} < n$)
or send a pair of names along its first parameter $b_1$ providing $L$ is not empty.
\begin{table}
\[ \begin{array}{l}
    F_{n, L}(b_1, b_2) \de \ip{b_2}{x_1, x_2}.F_{n, L_1}(b_1, b_2) \quad (\len{L} = 0) \\
    \qquad \text{where } L_1 = [(x_1, x_2)] \\

    F_{n, L}(b_1, b_2) \de \ip{b_2}{x_1, x_2}.F_{n, L_1}(b_1, b_2) + \op{b_1}{v_1, v_2}.F_{n, L_2}(b_1, b_2) \quad (0 < \len{L} < n) \\
    \qquad \text{where } L_1 = L \lc [(x_1, x_2)]; L = [(v_1, v_2)] \lc L_2 \\

    F_{n, L}(b_1, b_2) \de \op{b_1}{v_1, v_2}.F_{n, L_2}(b_1, b_2) \quad (\len{L} = n) \\
    \qquad \text{where } L = [(v_1, v_2)] \lc L_2
\end{array} \]
\caption{Buffer Process $F_{n, L}$}
\label{tab:pbbp}
\end{table}

The \emph{translation function} $\pbt{\cdot}$ takes a $\pi_b$ process and a valid buffer store as parameter
and returns a single $\pi$ process.
The encoding of a buffer store is a composition of buffer processes each representing a buffered name's buffer.
For processes, the encoding differs from the original process in the new operators and prefixes.
A new operator is encoded into two new operators localizing the pair of translation names.
The encoding of input prefix $\ip{c}{x}$ is also an input prefix
but the subject is $c$'s input name $c_1$,
while the encoding of output prefix $\op{c}{d}$ has the output name $c_2$ as the subject.
Finally, in the encoding of an extended new process $(\nu b:n)P$, a buffer process representing $b$'s buffer is added.

The \emph{action translation function} $\at$ maps $\pi_b$ actions to corresponding $\pi$ actions,
it is defined similar to the encoding of prefixes.

With an abuse of notation, we also write $\pbt{P}$ and $\pbt{\bs}$ for the encoding of $P$ and $\bs$ respectively.
The translation function $\pbt{\cdot}$, along with the bijective action translation function $\at$ are defined in Table~\ref{tab:pbt}.

\begin{table}
  \[ \begin{array}{lll}
    \pbt{\ip{c}{x}.P} & = & \ip{c_1}{x_1, x_2}.\pbt{P}  \\
    \pbt{\op{c}{d}.P} & = & \op{c_2}{d_1, d_2}.\pbt{P}  \\
    \pbt{\sum_{i \in I} \pi_i.P_i} & = & \sum_{i \in I} \pbt{\pi_i.P_i}  \\
    \pbt{P \pc Q} & = & \pbt{P} \pc \pbt{Q}  \\
    \pbt{(\nu b:n).P} & = & (\nu b_1 b_2)\tau.(\pbt{P} | F_{n, [\ ]}(b_1, b_2))  \\
    \pbt{(\nu c)P} & = & (\nu c_1 c_2)\pbt{P}  \\
    \pbt{!P} & = & !\pbt{P} \\
    \\
    \pbt{\bs} & = & \prod_{b \in \dom{\bs}} F_{n_b, L_b}(b_1, b_2)  \\
        &&\text{where } \bs(b) = (n_b, [d^1 \dots d^m]); L_b = [(d^1_1, d^1_2) \dots (d^m_1, d^m_2)] \\
    \\
    \pbt{P, \bs} &=& \left\{ \begin{array}{ll}
     (\nu c_1 c_2)\pbt{P', \bs\subs{c}{\nu c}} & \text{if } c \in ln(\bs) \cup (\dom{\bs} \cap ln(P)) \\
     & \text{and } P \equiv_\bs (\nu c)P' \\
     \pbt{P, \bs} = \pbt{P} \pc \pbt{\bs} & \text{otherwise}
    \end{array} \right.\\
    \\
    \at(\alpha) &=& \left\{
      \begin{array}{ll}
	    \ip{a_1}{d_1, d_2}         & \alpha = \ip{a}{d} \ \, \wedge a \not\in \dom{\bs} \\
	    \op{a_2}{d_1, d_2}         & \alpha = \op{a}{d} \ \, \wedge a \not\in \dom{\bs} \\
        \op{a_2}{\nu d_1, \nu d_2} & \alpha = \op{a}{\nu d} \wedge a \not\in \dom{\bs} \\
        \ip{b_2}{d_1, d_2}         & \alpha = \ip{b}{d} \ \, \wedge b \in \dom{\bs} \\
	    \op{b_1}{d_1, d_2}         & \alpha = \op{b}{d} \ \, \wedge b \in \dom{\bs} \\
        \op{b_1}{\nu d_1, \nu d_2} & \alpha = \op{b}{\nu d} \wedge b \in \dom{\bs} \\
	    \tau & \alpha = \tau \\
      \end{array} \right. \\
  \end{array} \]
  \caption{ Translation Function $\pbt{\cdot}$ and Action Translation Function $\at$}
  \label{tab:pbt}
\end{table}

The following properties are apparent.
Substitutions can be postponed until after the translation.
\begin{proposition}
  For a process $P$, and the translation function $\pbt{\cdot}$
  \[ \pbt{P\subs{c}{x}} = \pbt{P}\subs{c_1, c_2}{x_1, x_2} \]
  where $\nt(x) = (x_1, x_2)$ and $\nt(c) = (c_1, c_2)$
  \label{lem:subs}
\end{proposition}

And structure congruent processes have the 'same' encodings.
\begin{proposition}
  If $P \equiv_{\bs} Q$, then $\pbt{P, \bs} \equiv \pbt{Q, \bs}$
  \label{lem:sctrans}
\end{proposition}

These propositions can by proved by induction on the structure of $P$

\subsubsection{Full Abstraction}

The following two lemmas show that transitions of a $\pi_b$ process can be simulated by its encoding,
and no more transition is introduced by the encoding.

\begin{lemma}
  Suppose $(P, \bs) \xra{\alpha} (P', \bs')$, then $\pbt{P, \bs} \xra{\at(\alpha)} \pbt{P', \bs'}$.
  \label{lem:gosim1}
\end{lemma}

\begin{lemma}
  Suppose $\pbt{P, \bs} \xra{\at(\alpha)} R$, then $(P, \bs) \xra{\alpha} (P', \bs')$ and $R = \pbt{P', \bs'}$.
  \label{lem:gosim2}
\end{lemma}

Strong bisimulation relation is retained in the translation.

\begin{lemma}
  If $(P, \bs_P) \dotsim (Q, \bs_Q)$, then $\pbt{P, \bs_P} \dotsim \pbt{Q, \bs_Q}$.
  \label{lem:complete}
\end{lemma}
\begin{proof}
  We show the following relation $\mathcal{R}$ is a strong bisimulation.
  \[ \mathcal{R} = \Big\{ (\pbt{P, \bs_P}, \pbt{Q, \bs_Q}) \ \Big| \ (P, \bs_P) \dotsim (Q, \bs_Q) \Big\} \]
  Suppose $\pbt{P, \bs_P} \xra{\at(\alpha)} R $, then by lemma~\ref{lem:gosim2}
  \[ R = \pbt{P', \bs'_P} \text{ and } (P, \bs_P) \xra{\alpha} (P', \bs'_P) \]
  Since $ (Q, \bs_Q) \xra{\alpha} (Q', \bs'_Q) $, by lemme~\ref{lem:gosim1}
  \[ \pbt{Q, \bs_Q} \xra{\at(\alpha)} \pbt{Q', \bs'_Q} \]
  and also $(\pbt{P', \bs'_P}, \pbt{Q', \bs'_Q}) \in \mathcal{R}$ because $ (P', \bs'_P) \dotsim (Q', \bs'_Q) $

  The other direction is the same. \qed
\end{proof}

\begin{lemma}
  If $\pbt{P, \bs_P} \dotsim \pbt{Q, \bs_Q}$, then $(P, \bs_P) \dotsim (Q, \bs_Q)$,
  \label{lem:sound}
\end{lemma}
\begin{proof}
  We show the following relation $\mathcal{R}$ is a strong bisimulation.
  \[ \mathcal{R} = \Big\{ \big( (P, \bs_P), (Q, \bs_Q) \big) \ \Big| \ \pbt{P, \bs_P} \sim \pbt{Q, \bs_Q} \Big\} \]
  Suppose $(P, \bs_P) \xra{\alpha} (P', \bs'_P) $, then by lemma~\ref{lem:gosim1}
  \[ \pbt{P, \bs_P} \xra{\at(\alpha)} \pbt{P', \bs'_P} \]
  Since $ \pbt{Q, \bs_Q} \xra{\at(\alpha)} R $, by lemme~\ref{lem:gosim2}
  \[ R = \pbt{Q', \bs'_Q} \text{ and } (Q, \bs_Q) \xra{\alpha} (Q', \bs'_Q) \]
  and also $((P', \bs'_P), (Q', \bs'_Q)) \in \mathcal{R}$ because $ \pbt{P', \bs'_P} \dotsim \pbt{Q', \bs'_Q} $

  The other direction is the same. \qed
\end{proof}

It follows that the encoding preserves strong bisimulation.
\begin{theorem}
  $(P, \bs_P) \dotsim (Q, \bs_Q)$ if and only if $\pbt{P, \bs_P} \dotsim \pbt{Q, \bs_Q}$.
\end{theorem}


\section{The Go Programming language}
\label{sec:go}

The Go programming language is a general purpose language developed by Google to support
easy and rapid development of large distributed systems.
\footnote{Goole has claimed that Go is used in production now.
Specifically, the website of Go (\url{golang.org}) and the download site (\url{dl.google.com}) are written in Go.}
This relatively young language inherits many good qualities of its ancestor while at the same time
introduces dozens of innovations for efficient and effective programming.
One of the most fascinating innovations is the concurrency feature which extremely simplifies
the construction of concurrent applications.
This section presents a formal operational semantics of the (core) Go language and a fully abstract
encoding in the $\pi_b$-calculus.

The syntax of a core of Go is presented in Table~\ref{tab:gosyn}.
An online specification of Go can be found at its website~\cite{GoSpec2012}.

\begin{table}
Types:
\[ \gotype{t} ::= \kw{int} \pc \kw{chan}\ \gotype{t} \]
Expressions:
\[  \begin{array}{rll}
    e, e_1, e_2, \ldots ::= & x & \mbox{Variable}
        \\ \pc & n & \mbox{Integer}
        \\ \pc & ch & \mbox{Channel}
        \\ \pc & \kw{make}(\kw{chan}\ \gotype{t}, n) & \mbox{Channel creation}
        \\ \pc & \laop e & \mbox{Receiving}
\end{array} \]
Statements:
\[
\begin{array}{rll}
    s, s_1, s_2, \ldots ::= & x = e & \mbox{Assignment}
        \\ \pc & e_1 \laop e_2 & \mbox{Sending}
        \\ \pc & s_1; s_2 & \mbox{Sequential}
        \\ \pc & \kw{go}\ f(e_1 \dots e_n) & \mbox{Go-routine}
        \\ \pc & \kw{select}\ \{c_1 \dots c_n\} & \mbox{Selection}
\end{array}
\]

where
$ c_1, c_2, \ldots ::= \kw{case}\ x = \laop e: s \pc \kw{case}\ e_1 \laop e_2: s$
  \caption{Syntax of the (core) Go}
  \label{tab:gosyn}
\end{table}

The \emph{channel type}, coupled with the concept called \emph{Go-routine}, constitutes the core of
Go's concurrency system.
Channel types are of the form $\kw{chan}\ \gotype t$, where $\gotype t$ is called the \emph{element type}.
Channels ($ch$) are first-class values of this language,
and they  are created by the make expression $\kw{make}(\kw{chan}\ \gotype t, n)$,
where $\kw{chan}\ \gotype{t}$ specifies the channel type and the integer $n$ specifies the size of the channel buffer.
Notice that $n$ must be non-negative and if it is zero, the created channel will be a synchronous channel.

Go-routines are similar to OS threads but much cheaper.
A Go-routine is launched by the statement $\kw{go}\ f(v_1 \dots v_n)$.
The function body of $f$ will be executed in parallel with the program that executes the $\kw{go}$ statement.
When the function completes, this Go-routine terminates and its return value is discarded.

Communication among Go-routines is achieved by sending and receiving operations on channels.
Sending statement $ch \laop v$ sends $v$ to channel $ch$, while receiving $\laop ch$,
regarded as an expression in Go, receives a value from $ch$.
Communication via unbuffered channels are synchronous.
Buffered (non-zero sized) channels enable asynchronous communication.
Sending a value to a buffered channel can proceed as long as its buffer is not full
and receiving from a buffered channel can proceed as long as its buffer is not empty.

\kw{select} statements introduce non-deterministic choice, but their clauses refer to only communication operations.
A \kw{select} statement randomly selects a clause whose communication is ``ready'' (able to proceed),
completes the selected communication, then proceeds with the corresponding clause statement.

Without loss of generality, we stipulate that a Go program is a set of function declarations,
each of the form
\[ \kw{func}\ f(x_1 \dots x_n)\ \{s\} \]
A Go program must specify a main function, which we shall refer to as $f_{start}$ in the sequel, as the entry point ---
running a Go program is equivalent to executing $\kw{go}\ f_{start} (\ldots)$ with appropriate arguments.
For the sake of simplicity, we only consider function calls in go statements and we assume that
all functions do not return values and  their bodies contain no local variables other than function arguments.

\subsection{Operational Semantics}
The structural operational semantics of Go is defined by a two-level labelled transition system:
the local transition system specifies the execution of a single Go-routine in isolation,
and the global transition system describes the behavior of a running Go program.

We first define the evaluation of expressions.
An expression configuration is a triple $\cf{e, \sigma, \delta_c}$,
where $e$ is the expression to be evaluated,
$\sigma$ is the {\em local store} mapping local variables to values,
and $\delta_c$ is the {\em channel store} mapping channels to triples $(n, l, g)$,
where $n$ is the capacity of the channel's buffer, $l$ is a list of values in the channel buffer,
and $g$ is a tag indicating whether the channel is local ($0$) or global ($1$).
The transition rules between expression configurations $\letgx{\alpha}$ are defined in Table~\ref{tab:golocexp},
where actions can be either silent action $\tau$, or $\gl{r}(ch, v)$ denoting receive action.
We often omit $\tau$ from silent transitions.

\begin{table}
\[ \begin{array}{l}
    \tts{Var}\ \cf{x, \sigma, \delta_c} \letg \cf{\sigma(x), \sigma, \delta_c}
    \vs \\

    \tts{Mak}\ \cfrac{ ch \not\in \dom{\delta_c} }
    { \cf{\kw{make}(\kw{chan}\ \tau, n), \sigma, \delta_c} \letg \cf{ch, \sigma, \delta_c[ch \mt (n, [\ ], 0)]} }
    \vs \\

    \tts{RvE}\ \cfrac{\cf{e, \sigma, \delta_c} \letgx{\alpha} \cf{e', \sigma, \delta_c'}}
    { \cf{\laop e, \sigma, \delta_c} \letgx{\alpha} \cf{\laop e', \sigma, \delta_c'}}
    \vs \\

    \tts{RvU}\ \cfrac{ \delta_c(ch) = (0, [\ ], g) }
    { \cf{\laop ch, \sigma, \delta_c} \letgx{\gl{r}(ch, v) } \cf{v, \sigma, \delta_c} }
    \vs \\

    \tts{RvB}\ \cfrac{ \delta_c(ch) = (n, [v] \lc l, g) \quad n > 0 }
    { \cf{\laop ch, \sigma, \delta_c} \letg \cf{v, \sigma, \delta_c[ch \mt (n, l, g)]} }
\end{array} \]
\caption{Transition Rules for Expressions}
\label{tab:golocexp}
\end{table}

\tts{Var} retrieves the value of $x$ from local store $\sigma$.
\tts{Mak} creates a fresh local channel $ch$.
Other rules concern receiving from channels.
Once the channel expression is fully evaluated, the real receive begins following rules \tts{RvU} and \tts{RvB}.
The value received from an unbuffered channel is indicated in the label, while the value received
from a buffered channel is the ``oldest'' value of the channel's buffer.

The local transition system defines transition rules between local configurations.
A local configuration is a tuple $\cf{s, \sigma, \delta_c}$, where $s$ is the statement to be executed,
$\sigma$ is the {\em local store} and $\delta_c$ is the {\em channel store}.
Each Go-routine has its own local store, but the channel store is shared by all Go-routines of a running program.
The local transition relation $\lag$ is presented in Table~\ref{tab:goloc}.
Two additional actions can occur in local transition rules: $\gl{s}(ch, v)$ for message sending over channels
and $\gl{g}(f, v_1 \dots v_n)$ for Go-routine creation.

\begin{table}
\[ \begin{array}{l}
    \tts{Ass}\ \cf{x = v, \sigma, \delta_c} \lag \cf{\kw{nil}, \sigma[x \mapsto v], \delta_c} \vs \\
    \tts{AsE}\ \cfrac{ \cf{e, \sigma, \delta_c} \letgx{\alpha} \cf{e', \sigma, \delta_c'} }
    { \cf{x = e, \sigma, \delta_c} \lagx{\alpha} \cf{x = e', \sigma, \delta_c'} } \vs \\

    \tts{SdU}\ \cfrac{ \delta_c(ch) = (0, [\ ], g) }{ \cf{ch \laop v, \sigma, \delta_c} \lagx{\gl{s}(ch, v) } \cf{\kw{nil}, \sigma, \delta_c} } \vs \\
    \tts{SdB}\ \cfrac{ \delta_c(ch) = (n, l, g);\ n > 0;\ \len{l} < n }
    { \cf{ch \laop v, \sigma, \delta_c} \lag \cf{\kw{nil}, \sigma, \delta_c[ch \mt (n, l \lc [v], g)]} } \vs \\
    \tts{SdE1}\ \cfrac{ \cf{e_1, \sigma, \delta_c} \letgx{\alpha} \cf{e'_1, \sigma, \delta_c'} }
    { \cf{e_1 \laop e_2, \sigma, \delta_c} \lagx{\alpha} \cf{e'_1 \laop e_2, \sigma, \delta_c'} } \vs \\
    \tts{SdE2}\ \cfrac{\cf{e_2, \sigma, \delta_c} \letgx{\alpha} \cf{e'_2, \sigma, \delta_c'} }
    { \cf{ch \laop e_2, \sigma, \delta_c} \lagx{\alpha} \cf{ch \laop e'_2, \sigma, \delta_c'}} \vs \\

    \tts{Go}\quad \cf{\kw{go}\ f(v_1 \dots v_n), \sigma, \delta_c} \lagx{\gl{g}(f, v_1 \dots v_n)} \cf{\nl, \sigma, \delta_c} \vs \\
    \tts{GoE}\ \cfrac{ \cf{e_i, \sigma, \delta_c} \letgx{\alpha} \cf{e'_i, \sigma, \delta_c'} }
    { \cf{\kw{go}\ f(.. v_{i-1}, e_i ..), \sigma, \delta_c} \lagx{\alpha} \cf{\kw{go}\ f(.. v_{i-1}, e'_i ..), \sigma, \delta_c'} } \vs \\

    \tts{SlR}\ \cfrac{\cf{\laop ch, \sigma, \delta_c} \letgx{\alpha} \cf{v, \sigma, \delta_c'};\ c_i \equiv \kw{case}\ x = \laop ch : s }
    { \cf{\kw{select}\ \{\dots c_i \dots\}, \sigma, \delta_c} \lagx{\alpha} \cf{x = v; s, \sigma, \delta_c'} } \vs \\
    \tts{SlS}\ \cfrac{\cf{ch \laop v, \sigma, \delta_c} \lagx{\alpha} \cf{\kw{nil}, \sigma, \delta_c'};\ c_i \equiv \kw{case}\ ch \laop v : s }
    { \cf{\kw{select}\ \{\dots c_i \dots\}, \sigma, \delta_c} \lagx{\alpha} \cf{s, \sigma, \delta_c'} } \vs \\

    \tts{SlE}\ \cfrac{\cf{e, \sigma, \delta_c} \letgx{\alpha} \cf{e', \sigma, \delta_c'} }
    { \cf{\kw{select}\ \{ .. c_i ..\}, \sigma, \delta_c} \lagx{\alpha} \cf{\kw{select}\ \{.. c'_i .. \}, \sigma, \delta_c'} } \\
    \ \left(\begin{array}{l}
      \text{where $e$ is the first non-fully evaluated subexpression} \\
      \text{of communication operations in $c_1 \dots c_n$}
    \end{array} \right) \vs \\

    \tts{Seq1}\ \cfrac{\cf{s_1, \sigma, \delta_c} \lagx{\alpha} \cf{s'_1, \sigma', \delta_c'} }
    { \cf{s_1; s_2, \sigma, \delta_c} \lagx{\alpha} \cf{s'_1; s_2, \sigma', \delta_c'}} \vs \\
    \tts{Seq2}\ \cfrac{\cf{s_2, \sigma, \delta_c} \lagx{\alpha} \cf{s'_2, \sigma', \delta_c'} }
    { \cf{\kw{nil}; s_2, \sigma, \delta_c} \lagx{\alpha} \cf{s'_2, \sigma', \delta_c'}}
\end{array} \]
\caption{Local Transition Rules of Go}
\label{tab:goloc}
\end{table}

Subexpression evaluation in Go is strict and leftmost, and this evaluation strategy is specified by
\tts{AsE}, \tts{SdE1}, \tts{SdE2}, \tts{GoE} and \tts{SlE}.
For select statement, its subexpressions are those in its communication operations ---
the $e_1, e_2$ in $\kw{case}\ e_1 \laop e_2 : s$ and the $e$ in $\kw{case}\ x = \laop e : s$.

Rules \tts{SdU} and \tts{SdB} capture the behavior of sending over unbuffered and buffered channels respectively.
Sending a value $v$ over an unbuffered channel $ch$ carries a sending label $\gl{s}(ch, v)$,
while sending over buffered channels is silent and can proceed as long as the target channel buffer is not full.
The \tts{Go} rule says that a go statement does nothing locally and can always proceed with a transition
with the $\gl{g}$ label ---  the label is here simply for notifying the global configuration to
generate corresponding Go-routines.
\tts{Ass} assigns $v$ to variable $x$. \kw{Seq1} and \kw{Seq2} specify the sequential execution.
In \tts{SlR} and \tts{SlS}, the select statement picks the $i$-th clause.

Global transitions happen between global configurations which contain information of all running Go-routines.
A global configuration, denoted by $\Lambda, \Lambda_1 \ldots$, is defined as a tuple $\cf{\Gamma, \delta_c}$,
where $\Gamma$ is a multi-set of statement/local store pairs $(s, \sigma)$, of all running Go-routines,
and $\delta_c$ is the channel store.

A global transition takes the form
\[ \delta_f \vdash \cf{\Gamma_1, \delta_{c_1}} \gagx{\alpha} \cf{\Gamma_2, \delta_{c_2}} \]
where $\delta_f$ is a mapping from function names to function definitions.
A Go program will start from an initial configuration
$\cf{\{(s_{start}, \sigma_{start})\}, \delta_{init} }$, where $s_{start}$ is the body of the main function $start$,
$\sigma_{start}$ is the local store of $start$, and $\delta_{init}$ is the initial channel store.
The global transition rules are listed in Table~\ref{tab:goglo}.
A global action can be either $\tau$, $\gl{r}(ch, v)$ or $\gl{s}(ch, v)$.

\begin{table} {
\[ \begin{array}{l}
    \tts{Loc}\ \cfrac{ \cf{s, \sigma, \delta_c} \lag \cf{s', \sigma', \delta_c'} }
    { \delta_f \vdash \cf{\Gamma \cup \{(s, \sigma)\}, \delta_c} \gag \cf{\Gamma \cup \{(s', \sigma')\}, \delta_c'} }
    \vs \\

    \tts{Com}\ \cfrac{ \cf{s_1, \sigma_1, \delta_c} \lagx{\gl{r}(ch, v)} \cf{s'_1, \sigma_1, \delta_c};\
    \cf{s_2, \sigma_2, \delta_c} \lagx{\gl{s}(ch, v)} \cf{s'_2, \sigma_2, \delta_c} }
    { \delta_f \vdash \cf{\Gamma \cup \{(s_1, \sigma_1), (s_2, \sigma_2)\}, \delta_c} \gag \cf{\Gamma \cup \{(s'_1, \sigma_1), (s'_2, \sigma_2)\}, \delta_c} }
    \vs \\

    \tts{LGo}\ \cfrac{ \cf{s, \sigma, \delta_c} \lagx{\gl{g}(f, v_1 \dots v_m)} \cf{s', \sigma, \delta_c};\
    \delta_f(f) = (\kw{func}\ f(x_1 \dots x_m)\ \{s_f\}) }
    { \delta_f \vdash \cf{\Gamma \cup \{(s, \sigma)\}, \delta_c} \gag
    \cf{\Gamma \cup \{(s', \sigma), (s_f, [x_1 \mt v_1 \dots x_m \mt v_m])\}, \delta_c} }
    \vs \\

    \tts{GRU}\ \cfrac{ \cf{s, \sigma, \delta_c} \lagx{\gl{r}(ch, v)} \cf{s', \sigma, \delta_c};\ \delta_c(ch) = (0, [\ ], 1) }
    { \delta_f \vdash \cf{\Gamma \cup \{(s, \sigma)\}, \delta_c} \gagx{\gl{r}(ch, v)} \cf{\Gamma \cup \{(s', \sigma)\}, \delta_c} }
    \vs \\

    \tts{GRB}\ \cfrac{ \delta_c(ch) = (n, l, 1);\ n > 0;\ \len{l} < n }
    { \delta_f \vdash \cf{\Gamma, \delta_c} \gagx{\gl{r}(ch, v)} \cf{\Gamma, \delta_c[ch \mt (n, l \lc [v], 1)]} }
    \vs \\

    \tts{GSU1}\ \cfrac{ \cf{s, \sigma, \delta_c} \lagx{\gl{s}(ch, v)} \cf{s', \sigma, \delta_c};\ \delta_c(ch) = (0, [\ ], 1);\ v \not\in \dom{\delta_c} }
    { \delta_f \vdash \cf{\Gamma \cup \{(s, \sigma)\}, \delta_c} \gagx{\gl{s}(ch, v)} \cf{\Gamma \cup \{(s', \sigma)\}, \delta_c} }
    \vs \\

    \tts{GSU2}\ \cfrac{ \cf{s, \sigma, \delta_c} \lagx{\gl{s}(ch, ch')} \cf{s', \sigma, \delta_c};\ \delta_c(ch) = (0, [\ ], 1);\ \delta_c(ch') = (n', l', g') }
    { \delta_f \vdash \cf{\Gamma \cup \{(s, \sigma)\}, \delta_c} \gagx{\gl{s}(ch, \nu ch')} \cf{\Gamma \cup \{(s', \sigma)\}, \delta_c[ch' \mt (n', l', 1)]} }
    \vs \\

    \tts{GSB1}\ \cfrac{ \delta_c(ch) = (n, [v] \lc l, 1);\ n > 0;\ v \not\in \dom{\delta_c} }
    { \delta_f \vdash \cf{\Gamma, \delta_c} \gagx{\gl{s}(ch, v)} \cf{\Gamma, \delta_c[ch \mt (n, l, 1)]} }
    \vs \\

    \tts{GSB2}\ \cfrac{ \delta_c(ch) = (n, [ch'] \lc l, 1);\ n > 0;\ \delta_c(ch') = (n', l', g') }
    { \delta_f \vdash \cf{\Gamma, \delta_c} \gagx{\gl{s}(ch, \nu v)} \cf{\Gamma, \delta_c[ch \mt (n, l, 1), ch' \mt (n', l', 1)]} }
\end{array} \] }
\caption{Global Transition Rules of Go}
\label{tab:goglo}
\end{table}

\tts{Loc} specifies the independent transition of a single Go-routine.
Asynchronous communication will also take this transition since \tts{RvB} and \tts{SdB} are both silent transitions.
\tts{LGo} creates a new Go-routine.
\tts{Com} defines the synchronous communication between two Go-routines over unbuffered channels.
The rules \tts{Loc}, \tts{LGo} and \tts{Com} all specify internal actions of a running program.

A Go program can communicate with the environment via global channels.
\tts{GRU}, \tts{GSU1} and \tts{GSU2} describe how a Go program interact with the environment via unbuffered channels,
and \tts{GRB}, \tts{GSB1} and \tts{GSB2} describe interactions via buffered channels.
Because communication over buffered channels are asynchronous, the labels in \tts{GRB}, \tts{GSB1} and \tts{GSB2}
indicate how a global channel interacts with the environment.
For instance, in \tts{GRB} the label $\gl{r}(ch, v)$ means that the channel (buffer) $ch$ receives a value $v$
from the environment.
The two rules \tts{GSU2} and \tts{GSB2} also describe how a local channel is exposed to the environment and
becomes a global channel, by communication upon global channels.
The $\nu$ in the label is required only when the value is a local channel ($g' = 0$).

Let $t = \alpha_1 \dots \alpha_n$ where each $\alpha_i$ is a global action, we write $\hat{t}$ for the action sequence obtained by eliminating all the occurrences of $\tau$ in $t$. We write $P, \bs \gagx{t} P', \bs'$ if $P, \bs \gagx{\alpha_1} \cdots \gagx{\alpha_n} P', \bs'$, and $P, \bs \gagX{t} P', \bs'$ if $P, \bs \gagX{} \gagx{\alpha_1} \gagX{} \cdots \gagX{} \gagx{\alpha_n} \gagX{} P', \bs'$, where $\gagX{}$ is the reflexive and transitive closure of $\gagx{\tau}$.

\begin{definition}
  A symmetric binary relation $\mathcal{R}$ over global configurations is a \emph{(weak) bisimulation} if
  \[ \Lambda_1 \mathcal{R} \Lambda_2 \wedge \Lambda_1 \gagx{\alpha} \Lambda'_1 \text{ then } \exists \Lambda'_2\ .\ \Lambda_2 \gagX{\hat{\alpha}} \Lambda'_2 \wedge \Lambda'_1 \mathcal{R} \Lambda'_2 \]
  Two global configurations are bisimilar, written as $\Lambda_1 \approx_g \Lambda_2$, if they are related by some bisimulation.
\label{def:gcwb}
\end{definition}
Two Go programs $gp_1, gp_2$ are bisimilar, if their initial global configurations (with the same $\delta_c$) are bisimilar.

\subsection{Encoding}
The encoding of Go in the $\pi_b$-calculus is achieved by the translation function $\tsg{\cdot}(r)$,
which maps Go expressions and statements to $\pi_b$ processes.
The parameter $r$ is the name along which the result of an expression is returned or the termination of a statement is signaled.
The translation function $\tsg{\cdot}$ is defined in Table~\ref{tab:goenc}.

\begin{table}
\[ \begin{array}{l}
    \tts{VDec}\ Var(x, v) =  (\nu t)(\op{t}{v} | !t(z).\ip{x}{g, p}.(\ip{p}{y}.\op{t}{y} | \op{g}{z}.\op{t}{z})) \\ \\

    \tts{L2RE}\ LR(v_1 \dots v_n, r) = \op{r}{v_1 \dots v_n} \vs  \\
    \phantom{\tts{L2RE}\ } LR(\dots e^1, e^2 \dots e^m \dots, r) = (\nu r_1 t_2 r_2 \dots t_m r_m)( \vs  \\
     \qquad \qquad \qquad  \tsg{e^1}(r_1) | t_2.\tsg{e^2}(r_2) | \dots | t_m.\tsg{e^m}(r_m) \vs  \\
     \qquad \qquad \qquad |\ \ip{r_1}{v^1}.\ol{t_2}.\ip{r_2}{v^2} \dots \ol{t_m}.\ip{r_m}{v^m}.\op{r}{v_1 \dots v_n} \\ \\

    \tts{Valu}\ \tsg{n}(r) = \op{r}{n} \quad \tsg{ch}(r) = \op{r}{ch} \vs  \\

    \tts{Vari}\ \tsg{x}(r) = (\nu g p)(\op{x}{g, p}.\kt{g(z)}.\op{r}{z}) \vs  \\

    \tts{Recv}\ \tsg{\laop e}(r) = (\nu r')(\tsg{e}(r')|r'(y).\kt{\ip{y}{z}}.\op{r}{z}) \vs  \\

    \tts{Make}\ \tsg{\kw{make}(\kw{chan}\ \tau, 0)}(r) = \kt{\tau}.(\nu a)\op{r}{a} \vs  \\
    \phantom{\tts{Make}\ } \tsg{\kw{make}(\kw{chan}\ \tau, n)}(r) = \kt{(\nu b:n)}\op{r}{b} \\ \\

    \tts{Nil}\ \ \, \tsg{\nl}(r) = \ol{r} \vs  \\

    \tts{Assi}\ \tsg{x = e}(r) = (\nu r')(\tsg{e}(r')|r'(z).(\nu g p)(\op{x}{g, p}.\kt{\op{p}{z}}.\ol{r})) \vs  \\

    \tts{Send}\ \tsg{e_1 \laop e_2}(r) = (\nu r')(LR(e_1, e_2, r') | \ip{r'}{y, z}.\kt{\op{y}{z}}.\ol{r}) \vs  \\

    \tts{Sequ}\ \tsg{s_1;s_2}(r) = (\nu r')(\tsg{s_1}(r') | r'.\tsg{s_2}(r)) \vs  \\

    \tts{Go}\quad\ \tsg{\kw{go}\ f(e_1 \dots e_n)}(r) = (\nu r')(LR(e_1 \dots e_n, r') \vs  \\
    \qquad \qquad \qquad | \ip{r'}{y_1 \dots y_n}.\kt{\op{f}{y_1 \dots y_n}}.\ol{r}) \vs  \\

    \tts{Sele}\ \tsg{\kw{select}\ \{c_1 \dots c_n\}}(r) = (\nu r')(LR(e_1 \dots e_k, r') \vs  \\
    \qquad \qquad \qquad | \ip{r'}{y_1 \dots y_k}.(SC_{c_1} + \dots + SC_{c_n})) \vs  \\
    \qquad SC_c = \left\{ \begin{array}{ll}
      \kt{\ip{y_i}{v}}.\tsg{x = v; s}(r)\qquad & c \equiv \kw{case}\ x = \laop e_i : s \vs  \\
      \kt{\op{y_i}{y_j}}.\tsg{s}(r) & c \equiv \kw{case}\ e_i \laop e_j : s \vs  \\
    \end{array} \right. \\ \\
\end{array} \]
\caption{ Encoding of Go }
\label{tab:goenc}
\end{table}

In the encoding, we use synchronous communication via local names to arrange the evolution order of $\pi_b$ processes.
For instance, in \tts{Recv}, the right hand side of the composition will not proceed
unless the left hand side outputs along local name $r'$.

Process $Var(x, v)$ denotes variable $x$ whose current value is $v$.
After inputting a pair of local names $(g, p)$, one can retrieve the associated value by communicating on $g$
or update the variable by communicating on $p$.
Process $LR$ evaluates these non-fully evaluated expressions in an expression sequence in left-to-right order
by synchronous communication on local names.

\tts{Make} returns the local name denoting the newly created channel.
A receive operation corresponds to an input prefix in \tts{Recv}, while a send operation corresponds to an output prefix in \tts{Send}.
For the go statement, after evaluating the argument expressions, these arguments are sent to the function to which $f$ refers.
The statement does not wait for the function, rather it outputs the termination signal along $r$ immediately.

For select, suppose $e_1 \dots e_k$ is these (fully and non-fully evaluated) subexpressions appearing
in the communication operation of its clauses listed in lexical top-to-bottom and left-to-right order.
The encoding first evaluates this expression sequence, followed by a guarded choice each of its constituent denotes a select clause.
The use of guarded choice here seems unavoidable.

In the encoding, some prefixes and extended new operators are underlined.
They are the most significant part and will be discussed later.

The translation function can be extended to a mapping from global configurations (with $\delta_f$) to $\pi_b$ processes.
We write $\tsg{\Lambda}$ for the pair $(P, \bs)$, where $P$ is the encoding of $\Lambda$ and $\delta_f$, and $\bs$ is a valid buffer store inferred from channel store $\delta_c$.
The extended translation function is shown in Table~\ref{tab:goenc2}.

\begin{table}
\[ \begin{array}{l}
    \tts{FDec}\ \tsg{\kw{func}\ f(x_1 \dots x_n) \ \{s\}} = !f(z_1 \dots z_n).(\nu x_1 \dots x_n)( \vs \\
    \qquad \qquad \qquad Var(x_1, z_1) | \dots | Var(x_n, z_n) | (\nu r')\tsg{s}(r')) \\ \\

    \tts{LSto}\ \tsg{\sigma} = \prod_{i=1}^{n} Var(x_i, \sigma(x_i)) \quad x_i \in \dom{\sigma}; n = \len{\dom{\sigma}} \vs \\
    \tts{FEnv}\ \tsg{\delta_f} = \prod_{i=1}^{n} \tsg{\delta_f(f_i)} \qquad\quad\, f_i \in \dom{\delta_f}; n = \len{\dom{\delta_f}} \vs \\
    \tts{GRot}\ R_{s, \sigma} = \tsg{s, \sigma} = (\nu \tilde{x})( (\nu r)\tsg{s}(r) \pc \tsg{\sigma}) \vs \\
    \tts{GCon}\ \tsg{\Lambda} = \tsg{\cf{\{(s_1, \sigma_1) \dots (s_n, \sigma_n)\}, \delta_c}, \delta_f} \\
    \qquad \qquad \ \  = (\nu \tilde{ch})(\nu \tilde{f})( R_{s_1, \sigma_1} | \dots | R_{s_n, \sigma_n} | \tsg{\delta_f}), \dot{\delta_c}
\end{array} \]
\caption{Extended $\tsg{\cdot}$ }
\label{tab:goenc2}
\end{table}

The encoding of a function declaration is a replication of input prefix process.
Each replica starts by inputting the argument lists along $f$, followed by a composition of processes
denoting function parameters and function body.
Since function does not return anything and a normal function call is forbidden, the termination signal
is worthless, therefore a local name $r'$ is used in the encoding of the function body.

$R_{s, \sigma}$ represents a Go-routine in which $s$ is to be executed with local store $\sigma$.
Each $x$ in $\tilde{x}$ refers to a local variable.
In the encoding of a global configuration, the names referring to local channels ($\tilde{ch}$)
and functions ($\tilde{f}$) are local names.

The valid buffer store $\dot{\delta_c}$ is obtained from the channel store $\delta_c$ in three steps:
Firstly, prefix a $\nu$ symbol to the names referring to local channels in all buffers;
Secondly, remove these unbuffered channels from the domain of $\delta_c$;
And finally, for any buffered channel $ch$ in the domain of $\delta_c$, eliminate the third element $g$ from $\delta_c(ch)$.

\subsection{Correctness}

The correctness of the encoding is demonstrated by a full abstraction theorem with respect to (weak) bisimulation.
The following lemma says that a global transition may be simulated by a nontrivial sequence of transitions of its encoding.
Usually, the encoding will perform some internal adjustments before and after the real simulation.
\begin{lemma}
  If $\Lambda \gagx{\alpha} \Lambda'$, then $\tsg{\Lambda} \Ra \xra{\alpha'} \Ra \tsg{\Lambda'} $
  \label{lem:gocomp}
\end{lemma}
The lemma is proved by detailed analyze of the global transition rules one by one.

\begin{proposition}
  Suppose the transition is inferred by global transition rule \tts{Loc}, that is
  \[ \cfrac{ \cf{s, \sigma, \delta_c} \lag \cf{s', \sigma', \delta_c'} }
    { \delta_f \vdash \cf{\Gamma \cup \{(s, \sigma)\}, \delta_c} \gag \cf{\Gamma \cup \{(s', \sigma')\}, \delta_c'} }
    \]
  then
  \[ (\nu \tilde{ch}) R_{s, \sigma}, \dot{\delta_c} \Ra \ra \Ra (\nu \tilde{ch}) R_{s', \sigma'}, \dot{\delta'_c} \]
  where $\tilde{ch}$ are local channels.
\end{proposition}
\begin{proof}
Consider the local transition rules which can be applied in the last step of the inference of premise.

Suppose the premise is an instance of local transition rule \tts{Ass} or \tts{SdB},
the results follows by a detailed analyze on the actions of the encoding.

Suppose the premise is an instance of \tts{AsE}, and the premise of this instance is an instance
of local expression rule \tts{Var}, \tts{Mak} or \tts{RvB}, the results follows by a detailed analyze.
If the premise of this instance is an instance of local expression rule \tts{RvE},
we prove by induction on the depth of the inference of the premise of the instance.
Suppose $s \equiv x = \laop e, s' \equiv x = \laop e'$ where $\cf{e, \sigma, \delta'_c} \letg \cf{e', \sigma, \delta'_c}$ by a shorter inference.
By induction
\[ (\nu \tilde{ch})R_{x = e, \sigma}, \dot{\delta_c} \Ra \ra \Ra (\nu \tilde{ch})R_{x = e', \sigma}, \dot{\delta'_c} \]
From the definition of encoding for $x = e$ and $\laop e$, it follows that
\[ (\nu \tilde{ch})R_{x = \laop e, \sigma}, \dot{\delta_c} \Ra \ra \Ra (\nu \tilde{ch})R_{x = \laop e', \sigma}, \dot{\delta'_c} \]
\tts{SdE1}, \tts{SdE2}, \tts{GoE}, \tts{SlR}, or \tts{SlE} are similar to \tts{AsE}.

Suppose the premise is an instance of \tts{SlS}, \tts{SeQ1} or \tts{SeQ2},
then we prove by induction on the depth of the inference of the premise.
For \tts{SeQ1} $s \equiv s_1;s_2, s' \equiv s'_1;s_2$ where
$\cf{s_1, \sigma, \delta_c} \lag \cf{s'_1, \sigma', \delta'_c}$ by a shorter inference. By induction
\[ (\nu \tilde{ch})R_{s_1, \sigma}, \dot{\delta_c} \Ra \ra \Ra (\nu \tilde{ch})R_{s'_1, \sigma'}, \dot{\delta'_c} \]
From the definition of encoding for $s_1; s_2$, it follows that
\[ (\nu \tilde{ch})R_{s, \sigma}, \dot{\delta_c} \Ra \ra \Ra (\nu \tilde{ch})R_{s', \sigma'}, \dot{\delta'_c} \]
\tts{SlS} and \tts{SeQ2} are similar to \tts{SeQ1}.

  This completes the proof. \qed
\end{proof}

\begin{proposition}
  Suppose the transition is inferred by global transition rule \tts{LGo}, that is
  \[ \cfrac{ \cf{s, \sigma, \delta_c} \lagx{\gl{g}(f, v_1 \dots v_m)} \cf{s', \sigma, \delta_c} }
    { \delta_f \vdash \cf{\Gamma \cup \{(s, \sigma)\}, \delta_c} \gag \cf{\Gamma \cup \{(s', \sigma), (s_f, \sigma_f)\}, \delta_c} }
    \]
  where $\delta_f(f) = (\kw{func}\ f(x_1 \dots x_m)\ \{s_f\})$ and $\sigma_f = [x_1 \mt v_1 \dots x_m \mt v_m]$, then
  \[ (\nu \tilde{ch})(R_{s, \sigma} | \tsg{\delta_f}), \dot{\delta_c} \Ra \ra \Ra (\nu \tilde{ch})(R_{s', \sigma} | R_{s_f, \sigma_f} | \tsg{\delta_f} ), \dot{\delta'_c} \]
  where $\tilde{ch}$ are local channels.
\end{proposition}
\begin{proof}
Consider the local transition rules which can be applied in the last step of the inference of premise.

Suppose the premise is an instance of local transition rule \tts{Go},
the results follows by a detailed analyze on the actions of the encoding.

Suppose the premise is an instance of \tts{SeQ1} or \tts{SeQ2},
then we prove by induction on the depth of the inference of the premise.
For \tts{SeQ1} $s \equiv s_1;s_2, s' \equiv s'_1;s_2$ where
$\cf{s_1, \sigma, \delta_c} \lagx{\gl{g}(f, v_1 \dots v_m)} \cf{s'_1, \sigma, \delta_c}$ by a shorter inference. By induction
\[ (\nu \tilde{ch})(R_{s_1, \sigma} | \tsg{\delta_f}), \dot{\delta_c} \Ra \ra \Ra (\nu \tilde{ch})(R_{s'_1, \sigma} | R_{s_f, \sigma_f} | \tsg{\delta_f} ), \dot{\delta'_c} \]
From the definition of encoding for $s_1; s_2$, it follows that
\[ (\nu \tilde{ch})(R_{s_1;s_2, \sigma} | \tsg{\delta_f}), \dot{\delta_c} \Ra \ra \Ra (\nu \tilde{ch})(R_{s'_1;s_2, \sigma} | R_{s_f, \sigma_f} | \tsg{\delta_f} ), \dot{\delta'_c} \]
\tts{SeQ2} are similar to \tts{SeQ1}.

This completes the proof. \qed
\end{proof}

For other global transition, it is similar.

Conversely, a sequence of transitions of $\tsg{\Lambda}$ should reflect certain global transitions of $\Lambda$.
However it is not always possible, since the simulation may not yet complete, even worse
the transition sequence simulating one global transition may interleave with transition sequences simulating others.
Fortunately, by observing the proof of the previous lemma, we find that actually only one transition in the sequence
plays the crucial role, as this transition uniquely identifies a global transition.
Other $\tau$ transitions, whether preceding or following this special transition, are internal adjustments
which prepare for the special transition immediately after them.
We call the special transition a \emph{simulating transition}, and the other non-special $\tau$ transitions \emph{preparing transitions}.

Preparing transitions are local synchronous communication between subprocesses of one single Go-routine
(e.g. synchronous communication making subprocesses evolve in order).
To postpone or to advance preparing transitions would not affect the behavior of other Go-routines.

These observations are formulated by the following definitions and lemmas.

\begin{definition}
  A transition $P, B \xra{\alpha} P', B'$ is a simulating transition if the action $\alpha$ is induced by
  the underlined prefixes and extended new operators specified in the encoding in Table~\ref{tab:goenc}.
  Otherwise, it is a preparing transition.
  \label{def:gosim}
\end{definition}

\begin{definition}
  Let $\Lambda$ be a global configuration, the set $\tsgs{\Lambda}$ is defined as follows:
  \begin{enumerate} \setlength{\itemsep}{0ex}
    \item $\tsg{\Lambda} \in \tsgs{\Lambda} $
    \item If $(P, \bs) \in \tsgs{\Lambda}$ and $(P, \bs) \ra (P', \bs)$ is a preparing transition, then $(P', \bs) \in \tsgs{\Lambda}$
    \item If $(P, \bs) \in \tsgs{\Lambda}$ and $(P', \bs) \ra (P, \bs)$ is a preparing transition, then $(P', \bs) \in \tsgs{\Lambda}$
  \end{enumerate}
  \label{def:gotsset}
\end{definition}

\begin{proposition}
  Let $\Lambda$ be a global configuration, and $\tsg{\Lambda} = (\nu \tilde{ch})(\nu \tilde{f}) \\ (R_1 | \dots | R_n | \tsg{\delta_f}), \dot{\delta_c}$. Suppose $\tsg{\Lambda} \xRa{t} P, \bs$, then
  \[ P \equiv_\bs (\nu \tilde{ch'})(\nu \tilde{ch''})(\nu \tilde{f})(P_1| \dots | P_n | \dots | P_m | \tsg{\delta_f}) \]
  where $P_i$ ($i \le n$) is (subprocess of) a descendant of $R_i$, $P_j$ ($j > n$) corresponds to a newly created Go-routine.
  Also, $\{\tilde{ch'}\} \subset \{\tilde{ch}\}$ and $\forall ch \in \{\tilde{ch}\} \setminus \{\tilde{ch'}\}\ .\ \op{d}{\nu ch} \in t$,
  where $d$ refers to a global channel when $\xra{\op{d}{\nu ch}}$ happens.
  Each name in $\tilde{ch''}$ denotes newly created channels during $\xRa{t}$.
  \label{prop:gosimstru}
\end{proposition}

\begin{proposition}
  If $P, \bs \ra P', \bs'$ is a preparing transition, then
  \begin{enumerate} \setlength{\itemsep}{0ex}
    \item $B$ = $\bs'$
    \item It is a preparing transition of $i$-th Go-routine, i.e. $P$ and $P'$ differs only on $P_i$ for some $i$.
    \item The transition is induced by $ P_i, \bs'' \ra P'_i, \bs'' $ where $\bs'' = \bs\subs{\tilde{ch}}{\nu \tilde{ch}}$.
  \end{enumerate}
  \label{prop:goprepare}
\end{proposition}

\begin{lemma}
  If $\tsgs{\Lambda} \ni (P, B) \Ra \xra{\alpha'} (P', B')$, and only $\xra{\alpha'}$ is a simulating transition, then there exists $\Lambda'$ such that
  \[ \Lambda \gagx{\alpha} \Lambda' \text{ and } (P', B') \in \tsgs{\Lambda'} \]
  \label{lem:gosound}
\end{lemma}
\begin{proof}
  By the Definition~\ref{def:gosim} and Propositions~\ref{prop:gosimstru} and \ref{prop:goprepare}. \qed
\end{proof}

Any of the processes in $\tsgs{\Lambda}$ can be seen as the encoding of $\Lambda$.
\begin{lemma}
  If $(P, B) \in \tsgs{\Lambda}$ and $(Q, B) \in \tsgs{\Lambda}$, then we have $(P, B) \approx (Q, B)$.
  \label{lem:gotsgs}
\end{lemma}
\begin{proof}
  The following relation is a bisimulation.
  \[ \mathcal{R} = \{ ((P, \bs), (Q, \bs)) \ | \ (P, \bs) \in \tsgs{\Lambda} \wedge (Q, \bs) \in \tsgs{\Lambda} \}\]

  $(P, \bs) \ra (P', \bs') \in \tsgs{\Lambda}$ is trivial.

  Suppose $(P, \bs) \xra{\alpha'} (P', \bs') \not\in \tsgs{\Lambda}$ is a simulating transition. By Lemma~\ref{lem:gosound}, there exists $\Lambda'$ such that
  \[ \Lambda \gagx{\alpha} \Lambda' \text{ and } (P', \bs') \in \tsgs{\Lambda'} \]
  Suppose this $\gagx{\alpha}$ involves the $i$-th (and $j$-th) Go-routine. For $(Q, \bs)$, perform the preparing transitions of the $i$-th (and $j$-th) Go-routine, followed by $\xra{\alpha'}$, we have
  \[ (Q, \bs) \Ra\xra{\alpha'} (Q', \bs') \in \tsgs{\Lambda'} \]

  The other direction is similar. \qed
\end{proof}

As a consequence, bisimulation is preserved by the encoding.
\begin{theorem}
  $\Lambda_1 \approx_g \Lambda_2$ if and only if $\tsg{\Lambda_1} \approx \tsg{\Lambda_2}$.
  \label{thm:gofull}
\end{theorem}
\begin{proof}
  $\Rightarrow$: The following relation is a bisimulation up to $\approx$.
  \[ \mathcal{R} = \{  (\tsg{\Lambda_1}, \tsg{\Lambda_2}) \ | \  \Lambda_1 \approx_g \Lambda_2 \}\]
  Suppose $\tsg{\Lambda_1} \xra{\alpha'} (P, \bs_P) \not\in \tsgs{\Lambda_1}$ is a simulating transition. By Lemma~\ref{lem:gosound} and \ref{lem:gotsgs}, there exists $\Lambda'_1$ such that
  \[ \Lambda_1 \gagx{\alpha} \Lambda'_1 \ \text{and} \ \tsgs{\Lambda'_1} \ni (P, \bs_P) \approx \tsg{\Lambda'_1}  \]
  Since $\Lambda_1 \approx_g \Lambda_2$, there exist $\Lambda'_2$ such that
  \[ \Lambda_2 \gagX{\hat{\alpha}} \Lambda'_2 \approx_g \Lambda'_1 \]
  By Lemma~\ref{lem:gocomp}
  \[ \tsg{\Lambda_2} \xRa{\hat{\alpha'}} \tsg{\Lambda'_2} \in \tsgs{\Lambda'_2} \]
  The other direction is similar.

  $\Leftarrow$: The following relation is a bisimulation.
  \[ \mathcal{R} = \{ (\Lambda_1, \Lambda_2) \ | \ \tsg{\Lambda_1} \approx \tsg{\Lambda_2} \} \]
  Suppose $\Lambda_1 \gagx{\alpha} \Lambda'_1$, by Lemma~\ref{lem:gocomp}
  \[ \tsg{\Lambda_1} \Ra\xra{\alpha'}\Ra \tsg{\Lambda'_1} \]
  Since $\tsg{\Lambda_1} \approx \tsg{\Lambda_2}$, there exist $(Q, B_Q)$ such that
  \[ \tsg{\Lambda_2} \xRa{\hat{\alpha'}} (Q, B_Q) \approx \tsg{\Lambda'_1} \]
  For each simulating transitions of $\xRa{\hat{\alpha'}}$, by Lemma~\ref{lem:gosound}, there exists $\Lambda'_2$, such that
  \[ \Lambda_2 \gagX{\hat{\alpha}} \Lambda'_2 \text{ and } \tsgs{\Lambda'_2} \ni (Q, B_Q) \approx  \tsg{\Lambda'_2} \]
  The other direction is similar. \qed
\end{proof}


\section{Core Erlang}
\label{sec:erlang}
We improve the translation mapping showed in \cite{Noll2005} by a fully abstract encoding in the $\pi_b$-calculus.

\subsection{Syntax of Core Erlang}
\label{sec:erlsyn}
The syntax of a subset of Core Erlang is presented in Table~\ref{tab:cesyn}.

\begin{table}
Functions:
\[  f ::= \kw{fun}\ (x_1 \dotsc x_n)\ \raop\ e \]
Expressions:
\[ \begin{array}{rll}
e, e_1, e_2, \ldots ::= & n & \mbox{Integer}
\\ \pc & id & \mbox{Process id}
\\ \pc & x & \mbox{Variable}
\\ \pc & \kw{let}\ x = e_1\ \kw{in}\ e_2 & \mbox{Let binding}
\\ \pc & \kw{apply}\ fn(e_1 \dots e_n) & \mbox{Function application}
\\ \pc & \kw{spawn}\ fn\ [e_1 \dots e_n] & \mbox{Spawn Erlang Process}
\\ \pc & e_1\ !\ e_2 & \mbox{Sending}
\\ \pc & \kw{receive}\ c_1 \dots c_n & \mbox{receiving}
\end{array} \]
where
$ c_1, c_2, \ldots ::= x\ \kw{when}\ e_1\ \raop\ e_2$
\caption{Syntax of Core Erlang}
\label{tab:cesyn}
\end{table}

\kw{let} binds values to variables, and functions are bound to function names by function definitions in the form
\[  fn = f \]

The counterpart of Go-routine in Erlang is the \emph{Erlang process}.
Each Erlang process is identified by an unique process $id$.
Moreover every Erlang process is associated with a mailbox which is an unbounded ordered sequence.
The Erlang process is created by the \emph{spawn} expression.
This expression acts almost the same as the go statement except that
it is an expression and takes the newly created Erlang process's $id$ as result.

Communication in Erlang is asynchronous.
Send expression $e_1\ !\ e_2$ appends message $e_2$, which is also the result of this expression,
to the mailbox of the Erlang process identified by $e_1$.
Receive operation is based on pattern matching.
The receive clause is deliberately simplified to ``$x\ \kw{when}\ e_1\ \raop\ e_2$'', where $x$ is an ``always march'' pattern.
Once a receive expression occurs, messages reside in the mailbox of the Erlang process evaluating this expression
are tried in first-to-last order against the clauses $c_1 \dots c_n$ from left to right.
For message $v$ and clause $c$, pattern marching results in $x$ binding to $v$ in $e_1$ and $e_2$.
If the guard expression $e_1$ evaluates to the Erlang atom \kw{'true'}, matching succeeds,
message $v$ is deleted (received) from the mailbox, and the result of the expression is the result of $e_2$.
Otherwise, the next clause will be tried by $v$.
If no more clause left for $v$, that is $v$ does not march any clause, then the next message
in mailbox will be used for marching, with $v$ remains in the mailbox.
Sometimes none of the existing messages matches any clause, in this case receive blocks until new message arrives.

Without loss of generality, we stipulate a Core Erlang program is a set of function definitions,
in which a function named $start$ is defined.
Running a program is equivalent to evaluate $\kw{spawn}\ \kw{start}\ [\ldots]$ with appropriate arguments.
For the sake of simplicity, we assume the function bodies contain no local variables other than function arguments.
Note that function name may appear only at the function position of spawn or apply expressions
in this subset language --- high-order is not considered.

\subsection{Operational Semantics}
The structural operational semantics of Core Erlang is also defined by a two-level labelled transition system:
the local transition system specifies the evaluation of a single Erlang process in isolation,
and the global transition system describes the behavior of a running Erlang program.

The local transition system defines transition rules between local configurations.
A local configuration is a tuple $\cf{e, \mb}$ where $e$ is the expression to be evaluated
by Erlang process whose mailbox is $\mb$.
The local transition rules, defined in Table~\ref{tab:celoc}, are of the form
\[ \delta_f, id \vdash \cf{e, \mb} \laex{\alpha} \cf{e', \mb'} \]
where $\delta_f$ is a mapping from function names to functions,
and $id$ identifies the Erlang process evaluating the expression.
Actions can be either silent action $\tau$, $\erl{sd}(id, v)$ denoting send action,
or $\erl{sp}(fn, v_1 \dots v_n)$ denoting Erlang process creation

\begin{table}
\[ \begin{array}{l}
    \tts{Let}\ \delta_f, id \vdash \cf{\kw{let}\ x = v \ \kw{in}\ e_2, \mb} \lae \cf{e_2\subs{v}{x}, \mb} \vs \\
    \tts{LtE}\ \cfrac{ \delta_f, id \vdash \cf{e_1, \mb} \laex{\alpha} \cf{e'_1, \mb'} }
    { \delta_f, id \vdash \cf{\kw{let}\ x = e_1 \ \kw{in}\ e_2, \mb} \laex{\alpha} \cf{\kw{let}\ x = e'_1 \ \kw{in}\ e_2, \mb'}} \vs \\

    \tts{App}\ \cfrac{ \delta_f(fn) = \kw{fun}\ (\tilde{x})\ \raop\ e }
    { \delta_f, id \vdash \cf{\kw{apply}\ fn(\tilde{v}), \mb} \lae \cf{e\subs{\tilde{v}}{\tilde{x}}, \mb}} \vs \\
    \tts{ApE}\ \cfrac{ \delta_f, id \vdash \cf{e_i, \mb} \laex{\alpha} \cf{e'_i, \mb'};\ i \in [n] }
    { \delta_f, id \vdash \cf{\kw{apply}\ fn(\cdot e_i \cdot), \mb} \laex{\alpha} \cf{\kw{apply}\ fn(\cdot e'_i \cdot), \mb'} }  \vs \\

    \tts{Spa}\ \cfrac{ id' \text{ is fresh} }
    { \delta_f, id \vdash \cf{\kw{spawn}\ fn\ [\tilde{v}], \mb} \laex{\erl{sp}(id', fn, \tilde{v})} \cf{id', \mb}} \vs \\
    \tts{SpE}\ \cfrac{ \delta_f, id \vdash \cf{e_i, \mb} \laex{\alpha} \cf{e'_i, \mb'};\ i \in [n] }
    { \delta_f, id \vdash \cf{\kw{spawn}\ fn\ [\cdot e_i \cdot], \mb} \laex{\alpha} \cf{\kw{spawn}\ fn\ [\cdot e'_i \cdot], \mb'} }  \vs \\

    \tts{Snd}\ \delta_f, id \vdash \cf{id'\ !\ v, \mb} \laex{\erl{sd}(id', v)} \cf{v, \mb} \vs \\
    \tts{SdE1}\ \cfrac{ \delta_f, id \vdash \cf{e_1, \mb} \laex{\alpha} \cf{e'_1, \mb'} }{ \delta_f, id \vdash \cf{e_1\ !\ e_2, \mb} \laex{\alpha} \cf{e'_1\ !\ e_2, \mb'} } \vs \\
    \tts{SdE2}\ \cfrac{ \delta_f, id \vdash \cf{e_2, \mb} \laex{\alpha} \cf{e'_2, \mb'} }{ \delta_f, id \vdash \cf{e_1\ !\ e_2, \mb} \laex{\alpha} \cf{e_1\ !\ e'_2, \mb'} } \vs \\
    \\
    \tts{Rcv}\ \cfrac{ \left( \begin{array}{l}
      \forall 0 < l < k \wedge 0 < j \le n \\
      \qquad .\ \delta_f, id \vdash \cf{e^j_1\subs{\mb[l]}{x^j}, \mb} \lae \cf{\kw{'false'}, \mb} \vs \\
      \forall 0 < j < i \\
      \qquad .\ \delta_f, id \vdash \cf{e^j_1\subs{\mb[k]}{x^j}, \mb} \lae \cf{\kw{'false'}, \mb} \vs \\
      \delta_f, id \vdash \cf{e^i_1\subs{\mb[k]}{x^i}, \mb} \lae \cf{\ \kw{'true'}, \mb}
    \end{array} \right) }
    { \delta_f, id \vdash \cf{\kw{receive}\ c_1 \dots c_n, \mb} \lae \cf{e^i_2\subs{\mb[k]}{x^i}, \mb/[k]}} \vs \\
    \qquad \left( \begin{array}{l}
      \text{where } c_j \equiv x^j\ \kw{when}\ e^j_1\ \raop\ e^j_2 \text{ and } \mb/[k] \\
      \text{means remove the $k$-th element from $\mb$}
    \end{array} \right)

\end{array} \]
\caption{Local Transition Rules of Core Erlang}
\label{tab:celoc}
\end{table}
Subexpression evaluation in Core Erlang is strict, however, in which order a sequence of subexpressions are evaluated is not defined.
This evaluation strategy is specified by \tts{LtE}, \tts{ApE}, \tts{SpE}, \tts{SdE1} and \tts{SdE2}.

\tts{Let} and \tts{App} is straightforward.
In \tts{Spa}, the $\erl{sp}$ label indicates that the new Erlang process is identified by $id'$
and the expression it will evaluate is the function application $\kw{apply}\ fn(\tilde{v})$.
Sending a message to an Erlang process carries the sending label $\erl{sd}(id, v)$, while receiving is silent.
The premise of \tts{Rcv} indicates that the first suitable message is the $k$-th message, and it marches the $i$-th clause.

Global transitions happen between global configurations which contain information of all running Erlang processes.
A global configuration, denoted by $\Lambda, \Lambda_1, \ldots$, is defined as a tuple $\cf{ID, E, \delta_m}$,
where $ID$ and $E$ are the sets of $id$s and expressions, respectively, of all running Erlang processes, and $\delta_m$ is the mailbox store.
A mailbox store $\delta_m$ is a mapping from process $id$s to pairs $(\mb, g)$, where $\mb$ is a mailbox (a list)
and $g$ is a tag indicating whether (the mailbox of) the Erlang process is accessible by an observer ($1$) or not ($0$).

We say an Erlang process is local if it is created during the evaluation of a program.
The set of local Erlang process $id$s is exactly the $ID$ of a global configuration.
All the Erlang processes in the environment (context) are global.
An local Erlang process is accessible if the environment knows its process $id$.

A global transition takes the form
\[ \delta_f \vdash \cf{ID_1, E_2, \delta_{m_1}} \gaex{\alpha} \cf{ID_2, E_2, \delta_{m_2}} \]
An Erlang program will start from an initial configuration
\[ \cf{\{id_{start}\}, \{e_{start}\}, \delta_{init} } \]
where $id_{start}$ is a fresh process $id$,
$e_{start}$ is the expression obtained from the body of $start$ by simultaneously substituting
supplied arguments for parameters of the function,
and $\delta_{init} = [id_{start} \mt ([\ ], 0)]$ is the initial mailbox store.
The global transition rules are listed in Table~\ref{tab:ceglo}.
A global action can be either $\tau$, $\gl{s}(id, v)$ for sending, or $\gl{r}(id, v)$ for receiving.

\begin{table}
\[ \begin{array}{l}
    \tts{Loc}\ \cfrac{ \delta_f, id \vdash \cf{e, \mb} \lae \cf{e', \mb'} }
    { \delta_f \vdash \cf{ID, \{\cdot e \cdot\}, \delta_m} \gae \cf{ID, \{\cdot e' \cdot\}, \delta_m[id \mt (\mb', g)]}}
    \vs \\

    \tts{LSp}\ \cfrac{ \delta_f, id \vdash \cf{e, \mb} \laex{\erl{sp}(id', fn, \tilde{v})} \cf{e'', \mb} ;\
            id' \not\in ID ;\
            \delta_f(fn) = \kw{fun}\ (\tilde{x})\ \raop\ e' }
    { \delta_f \vdash \cf{ID, \{\cdot e \cdot\}, \delta_m} \gae \cf{ID \cup \{id'\}, \{\cdot e'' \cdot\} \cup \{e'\subs{\tilde{v}}{\tilde{x}}\}, \delta_m[id' \mt ([\ ], 0)] } }
    \vs \\

    \tts{LSd}\ \cfrac{ \delta_f, id \vdash \cf{e, \mb} \laex{\erl{sd}(id', v)} \cf{e', \mb} ;\
        id' \in ID ;\
        \delta_m(id') = (\mb', g') }
    { \delta_f \vdash \cf{ID, \{\cdot e \cdot\}, \delta_m} \gae \cf{ID, \{\cdot e' \cdot\}, \delta_m[id' \mt (\mb' \lc [v], g')]} }
    \vs \\

    \tts{GSd1}\ \cfrac{ \delta_f, id \vdash \cf{e, \mb} \laex{\erl{sd}(id', v)} \cf{e', \mb} ;\
        id' \not\in ID ;\
        v \not\in ID }
    { \delta_f \vdash \cf{ID, \{\cdot e \cdot\}, \delta_m} \gaex{\erl{s}(id', v)} \cf{ID, \{\cdot e' \cdot\}, \delta_m} }
    \vs \\

    \tts{GSd2}\ \cfrac{ \delta_f, id \vdash \cf{e, \mb} \laex{\erl{sd}(id', id'')} \cf{e', \mb} ;\
        id' \not\in ID ;\
        id'' \in ID ;\
        \delta_m(id'') = (\mb', g') }
    { \delta_f \vdash \cf{ID, \{\cdot e \cdot\}, \delta_m} \gaex{\erl{s}(id', \nu id'')} \cf{ID, \{\cdot e' \cdot\}, \delta_m[id'' \mt (\mb', 1)]} }
    \vs \\

    \tts{GRv}\ \cfrac{ id \in ID ;\
     \delta_m(id) = (\mb, 1) }
    { \delta_f \vdash \cf{ID, E, \delta_m} \gaex{\erl{r}(id, v)} \cf{ID, E, \delta_m[id \mt (\mb \lc [v], 1)]} }
\end{array} \]
\label{tab:ceglo}
\caption{Global Transition Rules of Core Erlang}
\end{table}

\tts{Loc} specifies the independent evaluation of an Erlang process.
Receive operation will also take this transition since \tts{Rcv} is a silent local transition.
\tts{LSp} creates a new Erlang process.
\tts{LSd} defines the sending operation between two \emph{local} Erlang processes.
The rules \tts{Loc}, \tts{LSp} and \tts{LSd} all specify internal actions of a running program.

The labels in \tts{GSD1}, \tts{GSD2} and \tts{GRV} indicate how an Erlang program interacts with the environment.
An Erlang program can send values to the environment via global Erlang process $id$s,
this behavior is captured by \tts{GSD1} and \tts{GSD2}.
The latter also describe how an inaccessible Erlang process becomes an accessible one.
Note that the $\nu$ symbol in the label $\erl{s}(id', \nu id'')$ in \tts{GSd2} is required only when $id''$ denotes an unaccessible Erlang process, i.e., $g' = 0$.
The environment can also send values to an accessible Erlang program via its $id$s.
In \tts{GRv} the label $\erl{r}(id, v)$ actually means that the accessible Erlang process $id$ ``receives'' a value $v$
from the environment.

\begin{definition}
  A symmetric binary relation $\mathcal{R}$ over global configurations is a \emph{(weak) bisimulation} if
  \[ \Lambda_1 \mathcal{R} \Lambda_2 \wedge \Lambda_1 \gaex{\alpha} \Lambda'_1 \text{ then } \exists \Lambda'_2\ .\ \Lambda_2 \gaeX{\hat{\alpha}} \Lambda'_2 \wedge \Lambda'_1 \mathcal{R} \Lambda'_2 \]
  Two global configurations are bisimilar, written as $\Lambda_1 \approx_e \Lambda_2$, if they are related by some bisimulation.
\label{def:ecwb}
\end{definition}
Two Erlang programs $ep_1, ep_2$ are bisimilar, if their initial global configurations (with the same $\delta_m$) are bisimilar.

\subsection{Encoding of Core Erlang}
\label{app:ceenc}
The encoding of Core Erlang in $\pi_b$ calculus is achieved by the translation function $\tse{e}(a, p, r)$. This function, defined in Table~\ref{tab:ceenc}, takes three parameter:
the first parameter $a$ stands for the $id$ and the ``input port'' of the mailbox of the Erlang process evaluating $e$;
the ``output port'' of the mailbox is obtained from the second parameter $p$;
and the result of $e$ is returned along the last parameter $r$.

\begin{table}
\[ \begin{array}{l}
    \tts{AtLi}\ \tse{al}(a, p, r) = \op{r}{al}
    \vs \\

    \tts{Vari}\ \tse{x}(a, p, r) = \op{r}{x}
    \vs \\

    \tts{Let}\ \ \, \tse{\kw{let}\ x = e_1 \ \kw{in}\ e_2}(a, p, r) = (\nu r')(\tse{e_1}(a, p, r') | \kt{\ip{r'}{x}}.\tse{e_2}(a, p, r))
    \vs \\

    \tts{Appl}\ \tse{\kw{apply}\ fn(e_1 \dots e_n)}(a, p, r) = (\nu r_1 \dots r_n)( \tse{e_1}(a, p, r_1) | \dots | \tse{e_n}(a, p, r_n)
    \vs \\
    \qquad \qquad \qquad | \ip{r_1}{z_1} \dots \ip{r_n}{z_n}.\kt{\op{fn}{a, p, r, z_1 \dots z_n}})
    \vs \\

    \tts{Spaw}\ \tse{\kw{spawn}\ fn\ [e_1 \dots e_n]}(a, p, r) = (\nu r_1 \dots r_n)( \tse{e_1}(a, p, r_1) | \dots | \tse{e_n}(a, p, r_n)
    \vs \\
    \qquad \qquad \qquad | \ip{r_1}{z_1} \dots \ip{r_n}{z_n} .(\nu b':\infty)(\nu a' p' r')(
    \vs \\
    \qquad \qquad \qquad \qquad Cp(a', b')|\op{p'}{b'} |\kt{\op{fn}{a', p', r' , z_1 \dots z_n}}.\op{r}{a'} )
    \vs \\

    \tts{Send}\ \tse{e_1\ !\ e_2}(a, p, r) = (\nu r_1 r_2)(\tse{e_1}(a, p, r_1) | \tse{e_2}(a, p, r_2) | \ip{r_1}{y}.\ip{r_2}{z}.\kt{\op{y}{z}}.\op{r}{z})
    \vs \\

    \tts{Recv}\ \tse{\kw{receive}\ c_1 \dots c_n}(a, p, r) = \ip{p}{b}.(\nu b':\infty)(\nu t s_1 \dots s_{n+1})(
    \vs \\
    \qquad \qquad \qquad \ol{t} | !RH | !RC_{c_1}(s_1, s_2) | \dots | !RC_{c_n}(s_n, s_{n+1}) )
    \vs \\

    \qquad \begin{array}{l}
      RH = t.b(z).\op{s_1}{z} | \ip{s_{n+1}}{z}.\op{b'}{z}.\ol{t} \vs \\

      RC_c(s, s') =  \ip{s}{x}.(\nu r')(\tse{e_1}(a, p, r') | \kt{\ip{r'}{y}} \vs \\
      \qquad \qquad \quad .\kw{if}\ y = \kw{'true'}\ \kw{then}\ \tse{e_2}(a, p, r) | Cp(b, b') | \op{p}{b'} \  \kw{else}\ \op{s'}{x} )  \vs \\

      Cp(c, b) = (\nu t)(\ol{t} \pc !\ t.\ip{c}{z}.\op{b}{z}.\ol{t})
    \end{array}

\end{array} \]
\caption{Encoding of Core Erlang}
\label{tab:ceenc}
\end{table}

In \tts{Spaw}, the input port (also the process $id$) and the output port of the new Erlang process's mailbox is $a'$ and $b'$ respectively.
Result of the function application is returned via local name $r'$ and hence simply dropped.

\subsubsection{Receive}
We use the following algorithm to simulate one receive operation.
\begin{table}
  \begin{enumerate}
      \item Suppose the previous buffered name is $b$. Create a new buffered name $b'$.
      \item \label{algo:recv:2} Retrieve a message $v$ from $b$, pattern match $v$ against the first clause.
      \item \label{algo:recv:3} Pattern match $v$ against clause ($x\ \kw{when}\ e_1\ \raop\ e_2$), substitute $v$ for each free $x$ in $e_1$ and $e_2$. If $e_1$ evaluates to $\kw{'true'}$, goto~\ref{algo:recv:4}; otherwise pattern match $v$ against the next clause, goto~\ref{algo:recv:3}. If no clause remains, insert $v$ into $b'$ and goto~\ref{algo:recv:2}.
      \item \label{algo:recv:4} Set the previous buffer to $b'$, copy all remaining messages from $b$ to $b'$, evaluate $e_2$.
    \end{enumerate}
    \caption{Receive Algorithm}
\end{table}

The algorithm uses two buffered names for each receive --- a newly created buffered name
and the buffered name created by a previously receive.
From the viewpoint of a receiver, the latter is the output port of the mailbox from which messages are retrieved.
Once the receive operation succeeds, the former will become the output port of the mailbox.
The encoding of receive expression is basically the implementation of the algorithm in the $\pi_b$-calculus

In \tts{Recv}, the previous buffered name, say $b$, is saved in the second parameter of the translation function ($p$).
After creating a new buffered name $b'$, process $RH$ is triggered.
Receive handle process $RH$ fetches a message from mailbox $b$ and passes it to the first clause process for matching.
Clause process $RC_c(s, s')$ gets the message from its first parameter.
If guard expression evaluates to $\kw{'true'}$, matching succeeds.
The corresponding clause body process begins its evaluation with the previous buffered name changed to $b'$ and a copying process $Cp$ carries all remaining messages from the old mailbox to the new one.
Otherwise, the message is passed along the second parameter to the next clause process for matching.
If the message does not match the last clause, it is passed back to $RH$ which then
inserts the message to the new mailbox $b'$ and starts the matching of the next message.
In the encoding, $RC$ is guarded by its first parameter $s$,
and all $RC$s are chained together by local names $s_1 \dots s_{n+1}$.
A clause process cannot proceed unless the matching of the previous clause failed.

In general, we have the following proposition concerning the mailbox.
\begin{proposition}
A mailbox is explicitly modelled as follows
\[ \begin{array}{l}
    \MB(a, b^n) = Cp(a, b) | Cp(b, b^1) | \dots | Cp(b^{n-1}, b^n)
\end{array} \]
where $a$ and $b^n$ are input and output ports of the mailbox.
Each \emph{buffered} $b^j$ ($j > 1$) is created by one receive operation,
and the first \emph{buffered} $b$ is created by the spawn expression.
Send expressions insert messages into the mailbox via input port $a$,
while receive expressions retrieve messages from the mailbox via output port $b^n$.
\end{proposition}

\subsubsection{Configuration}
The translation function can be extended to a mapping from global configurations (with $\delta_f$) to $\pi_b$ processes.
We write $\tse{\Lambda}$ for the pair $(P, \bs)$, where $P$ is the encoding of $\Lambda$ and $\delta_f$,
and $\bs$ is a valid buffer store inferred from mailbox store $\delta_m$.
The extended translation function is shown in Table~\ref{tab:ceenc2}.

\begin{table}
\[ \begin{array}{l}
    \tts{FDef}\ \tse{fn = \kw{fun}\ (x_1 \dots x_n)\ \raop\ e} = !\ip{fn}{a, p, r, x_1 \dots x_n}.\tse{e}(a, p, r) \\ \\

    \tts{FEnv}\ \tse{\delta_f} = \prod_{i=1}^{n} \tse{\delta_f(f_i)} \qquad\quad\, f_i \in \dom{\delta_f}; n = \len{\dom{\delta_f}} \vs \\
    \tts{EPro}\ R_{e, \mb}(a) = (\nu p r)(\nu b b^1 \dots b^k)(\tse{e}(a, p, r) \pc \op{p}{b^k} \pc \MB(a, b^k)) \vs \\
    \tts{ECon}\ \tsg{\Lambda} = \tse{\cf{ID, E, \delta_m}, \delta_f} \vs \\
    \qquad \qquad \ \ = (\nu \tilde{a})(\nu \tilde{fn})( R_{e_1, \mb_1}(a_1) | \dots | R_{e_n, \mb_n}(a_n) | \tse{\delta_f}), \dot{\delta_m}
\end{array} \]
\caption{Extended $\tse{\cdot}$ }
\label{tab:ceenc2}
\end{table}

The encoding of a function definition is a replication of input prefix process.
Each replica starts by inputting the argument lists along $fn$, followed by the processes denoting the function body.

$R_{e, \mb}(a)$ represents an Erlang Process ($a$) ready to evaluate expression $e$ with mailbox $\mb$ whose input port is $a$.
In the encoding of a global configuration, the names referring to local Erlang Processes
which are not accessible ($\tilde{a}$) and functions ($\tilde{fn}$) are local names.

\subsection{Correctness}
The correctness of the encoding can be demonstrated by a similar analyze as Go.
The following lemma says substitution for free variables can be postponed to after the encoding.

\begin{lemma}
  \[ \tse{e\subs{x}{v}}(a, p, r) = (\tse{e}(a, p, r))\subs{x}{v} \]
  \label{lem:cedelaysubs}
\end{lemma}
\begin{proof}
  Simple induction on the structure of $e$.
\end{proof}

\begin{theorem}
  If $\delta_f \vdash \Lambda \gaex{\alpha} \Lambda'$, then $\tse{\Lambda} \Ra \xra{\alpha'} \Ra \tse{\Lambda'}$
  \label{thm:cesim}
\end{theorem}

\begin{proposition}
  Suppose the transition is inferred by global transition rule \tts{Loc}, that is
  \[ \cfrac{ \delta_f, id \vdash \cf{e, \mb} \lae \cf{e', \mb'} }
    { \delta_f \vdash \cf{ID, \{\cdot e \cdot\}, \delta_m} \gae \cf{ID, \{\cdot e' \cdot\}, \delta_m[id \mt (\mb', g)]}}
  \]
  then
  \[ (\nu \tilde{a})R_{e, \mb}(a), \dot{\delta_m} \Ra \ra \Ra (\nu \tilde{a})R_{e', \mb'}(a), \dot{\delta'_m} \]
  where $\tilde{a}$ are local non-accessible Erlang Processes, and $\delta'_m = \delta_m[id \mt (\mb', g)]$.
\end{proposition}
\begin{proof}
We prove by simultaneous induction on the depth of inference of the premises.
Consider the local transition rules applied in the last step of the inference of premise.

For \tts{Rcv}: Suppose $e \equiv \kw{receive}\ c_1 \dots c_n; e' \equiv e^i_2\subs{\mb[k]}{x^i}; \mb' = \mb/[k]$, where
\[ \begin{array}{l}
    \forall 0 < l < k \wedge 0 < j \le n .\ \delta_f, id \vdash \cf{e^j_1\subs{\mb[l]}{x^j}, \mb} \lae \cf{\kw{'false'}, \mb} \\
    \forall 0 < j < i .\ \delta_f, id \vdash \cf{e^j_1\subs{\mb[k]}{x^j}, \mb} \lae \cf{\kw{'false'}, \mb} \\
    \delta_f, id \vdash \cf{e^i_1\subs{\mb[k]}{x^i}, \mb} \lae \cf{\kw{'true'}, \mb}
  \end{array} \]
by a shorter inference. By induction
\[ \begin{array}{l}
    \forall 0 < l < k \wedge 0 < j \le n .\ (\nu \tilde{a})R_{e^j_1\subs{\mb[l]}{x^j}, \mb}(a), \bs \Ra \ra \Ra (\nu \tilde{a})R_{\kw{'false'}, \mb}(a), \bs \\
    \forall 0 < j < i .\ \ (\nu \tilde{a})R_{e^j_1\subs{\mb[k]}{x^j}, \mb}(a), \bs \Ra \ra \Ra (\nu \tilde{a})R_{\kw{'false'}, \mb}(a), \bs \\
    (\nu \tilde{a})R_{e^i_1\subs{\mb[k]}{x^i}, \mb}(a), \bs \Ra \ra \Ra (\nu \tilde{a})R_{\kw{'true'}, \mb}(a), \bs
  \end{array} \]
From the definition of encoding for $\kw{receive}\ c_1 \dots c_n$, it follows that
\[ \begin{array}{l}
    \qquad (\nu \tilde{a})R_{e, \mb}(a), \dot{\delta_m} \\
    = (\nu p r)(\nu b b^1 \dots b^k)(\tse{ \kw{receive}\ c_1 \dots c_n }(a, p, r) | \op{p}{b^k} | \MB(a, b^k)), \dot{\delta_m} \\
    \Ra (\nu p r)(\nu b b^1 \dots b^{k+1}) \\
    \qquad \qquad \qquad (\tse{e^i_2\subs{\mb[k]}{x^i}}(a, p, r) | \op{p}{b^{k+1}} | \MB(a, b^k) | Cp(b^k, b^{k+1})), \dot{\delta'_m} \\
    = R_{e', \mb'}(a), \dot{\delta'_m}
  \end{array} \]

Other cases are similar. \qed
\end{proof}

\begin{proposition}
  Suppose the transition is inferred by global transition rule \tts{LSp}, that is
  \[ \cfrac{ \delta_f, id \vdash \cf{e, \mb} \laex{\erl{sp}(id', fn, \tilde{v})} \cf{e'', \mb} ;\
            id' \not\in ID }
    { \delta_f \vdash \cf{ID, \{\cdot e \cdot\}, \delta_m} \gae \cf{ID \cup \{id'\}, \{\cdot e'' \cdot\} \cup \{e'\subs{\tilde{v}}{\tilde{x}}\}, \delta'_m } }
  \]
  where $\delta_f(fn) = \kw{fun}\ (\tilde{x})\ \raop\ e'$ and $\delta'_m = \delta_m[id' \mt ([\ ], 0)]$, then
  \[ (\nu \tilde{a})(R_{e, \mb}(a) | \tse{\delta_f}) , \dot{\delta_m} \Ra \ra \Ra (\nu \tilde{a} a'')(R_{e'', \mb}(a) | R_{e'\subs{\tilde{v}}{\tilde{x}}, [\ ]}(a'')| \tse{\delta_f}), \dot{\delta'_m} \]
\end{proposition}

For other global transition, it is similar.

Although argument evaluation is strict in Core Erlang, the evaluation order of a sequence of argument expressions is undefined.
In the encoding, besides the interleaving between transitions of many Erlang processes, interleaving also exists
inside one single Erlang process --- between the transitions simulating argument evaluation.
This interleaving is not serious, since except for receive and send, expression evaluation has no side effect,
the transitions of one argument process will not affect the behavior of others.
For receive, the input prefix $\ip{p}{b}$ also acts as a semaphore which prevents two receive operations run in parallel.
For send, it may only affect the behavior of receive. But according to the operational semantics,
receive expression would proceed unless a legal message is already in the mailbox.
Hence the interleaving of transition simulating send and receive expression can be rearranged in a non-interleaving way.

\begin{definition}
  A transition $P, B \xra{\alpha} P', B'$ is a simulating transition if the action $\alpha$
  is induced by underlined prefixes specified in Table~\ref{tab:ceenc}.
  Otherwise, it is a preparing transition.
  For receive, only the transition $\xra{\tau}$ induced by $\xra{\ip{r'}{\kw{'true'}}}$ is a simulating transition.
  \label{def:cekey}
\end{definition}

\begin{definition}
  Let $\Lambda$ be a global configuration, the set $\tses{\Lambda}$ is defined as follows:
  \begin{enumerate}
    \item $\tse{\Lambda} \in \tses{\Lambda} $
    \item If $(P, \bs) \in \tses{\Lambda}$ and $ (P, \bs) \ra (P', \bs')$ is a preparing transition, then $(P', \bs') \in \tses{\Lambda}$
  \end{enumerate}
  \label{def:cetsset}
\end{definition}

\begin{lemma}
  If $\tses{\Lambda} \ni (P, B) \Ra \xra{\alpha'} (P', B')$, and only $\xra{\alpha'}$ is a simulating transition,
  then there exists $\Lambda'$ such that
  \[ \Lambda \gaex{\alpha} \Lambda' \quad \text{and} \quad (P', B') \in \tses{\Lambda'} \]
  \label{lem:ceexist}
\end{lemma}

Any of the processes in $\tses{\Lambda}$ can be seen as the encoding of $\Lambda$.
\begin{lemma}
  If $(P, B) \in \tses{\Lambda}$ and $(Q, B') \in \tses{\Lambda}$, then we have $(P, B) \approx_e (Q, B')$
\end{lemma}

As a consequence, bisimulation is preserved by the encoding.
\begin{theorem}
  $ \Lambda_1 \approx_e \Lambda_2$ if and only if $\tse{\Lambda_1} \approx \tse{\Lambda_2} $
\end{theorem}


\section{Conclusion and Future Work}
\label{sec:con}

We have presented the $\pi_b$-calculus which extends the $\pi$-calculus by buffered names.
Communication along buffered names is asynchronous, i.e. native support of asynchronous communication.
After presenting its syntax and semantics, we give out a full abstract encoding of the $\pi_b$ calculus in the traditional poayadic $\pi$-calculus
with respect to strong bisimulation.
It is obvious that the new calculus does not increase the expressive power.
However, in contrast to the $\pi$-calculus which is hard to use in practice, it enables easy and clear modeling of practical concurrent languages.
Specifically, we have provided encodings of two real-world concurrent languages --- the (core) Go language and the Core Erlang --- in the buffered $\pi$-calculus.
Both encodings are fully abstract with respect to weak bisimulations.

The transition rules of the $\pi_b$-calculus are a bit complicated compared with that of the $\pi$-calculus.
We aim at applying the new language for modeling and verifying large distributed and concurrent systems
with asynchronous message passing-like communication by automatic computer programs.
One line of future work is to develop such programs.
We may extend existing tools such as Pict~\cite{Pierce2000}, MWB~\cite{Victor1994} or the HD Automata Laboratory~\cite{Ferrari2003}
to handle the $\pi_b$-Calculus.

Since weak bisimulation is not sufficient to demonstrate program equivalence, we may expect some full abstraction encodings
with respect to branching bisimulation, or even strong bisimulation.


\section*{Acknowledgement}
The authors are partially supported by Natural Science Foundation of China (61173033, 61033002, 61100053).
They would like to thank the members of BASICS for their interest in this work.
They are also grateful to the three anonymous referees for their detailed comments on the previous version of the paper.


\bibliographystyle{splncs03}
\bibliography{bpi}


\newpage

\appendix
\section{Proofs}

Proof of Lemma~\ref{lem:gosim1}

\begin{proof}
  We prove by induction on the depth of inference tree of the condition. Consider each rule in Table~\ref{tab:pbos}.

  \tts{IU}, \tts{OU}, \tts{IB}, \tts{OB}, \tts{IBG}, \tts{OBG} and \tts{NewB*} are the base step.

  \tts{IU} $P = \ip{a}{x}.P''; \alpha = \ip{a}{d}; P' = P''\subs{d}{x}; \bs' = \bs$
  \[ \pbt{P, \bs} = \ip{a_1}{x_1, x_2}.\pbt{P''} | \pbt{\bs} \xra{\ip{a_1}{d_1, d_2}} \pbt{P''}\subs{d_1, d_2}{x_1, x_2} | \pbt{\bs} = \pbt{P', \bs} \]

  \tts{OU} $P = \op{a}{d}.P'; \alpha = \op{a}{d}; \bs' = \bs$
  \[ \pbt{P, \bs} = \op{a_2}{d_1, d_2}.\pbt{P'} | \pbt{\bs} \xra{\op{a_2}{d_1, d_2}} \pbt{P'} | \pbt{\bs} = \pbt{P', \bs} \]

  \tts{IB} $P = \ip{b}{x}.P''; \alpha = \tau; P' = P''\subs{d}{x}; \bs(b) = (n, [d] \lc l); \bs' = \bs[b \mt (n, l)]$
  \[ \pbt{P, \bs} = \ip{b_1}{x_1, x_2}.\pbt{P''} | F_{n, [(d_1, d_2)] \lc L_l}(b_1, b_2) | \pbt{\bs \backslash b} \]
  \[ \pbt{P, \bs} \xra{\tau} \pbt{P''}\subs{d_1, d_2}{x_1, x_2} | F_{n, L_l}(b_1, b_2) | \pbt{\bs \backslash b} = \pbt{P', \bs'} \]

  \tts{OB} $P = \op{b}{d}.P'; \alpha = \tau; \bs(b) = (n, l); \bs' = \bs[b \mt (n, l \lc [d])] $
  \[ \pbt{P, \bs} = \op{b_2}{d_1, d_2}.\pbt{P'} | F_{n, L_l}(b_1, b_2) | \pbt{\bs \backslash b} \]
  \[ \pbt{P, \bs} \xra{\tau} \pbt{P'} \pc F_{n, L_l \lc [(d_1, d_2)]}(b_1, b_2) \pc \pbt{\bs \backslash b} = \pbt{P', \bs'} \]

  \tts{IBG} $P' = P; \alpha = \ip{b}{d}; \bs(b) = (n, l); \bs' = \bs[b \mt (n, l \lc [d])] $
  \[ \pbt{P, \bs} = (\nu \tilde{c_1} \tilde{c_2})( \pbt{Q} \pc \pbt{\bs\subs{\tilde{c}}{\nu \tilde{c}} \backslash b } \pc F_{n, L_l}(b_1, b_2) ) \]
  where $\{\tilde{c}\} = ln(\bs) \cup (\dom{\bs} \cap ln(P)), P \equiv_\bs (\nu \tilde{c})Q$.
  Since $b \in \dom{\bs}$ and $b \not\in ln(P)$, then $b \not\in \{\tilde{c}\}$
  \[ \pbt{P, \bs} \xra{\ip{b_2}{d_1, d_2}} (\nu \tilde{c_1} \tilde{c_2})( \pbt{Q} \pc \pbt{\bs\subs{\tilde{c}}{\nu \tilde{c}} \backslash b } \pc F_{n, L_l \lc [(d_1, d_2)]}(b_1, b_2) ) =  \pbt{P', \bs'} \]

  \tts{OBG} $P' = P; \alpha = \op{b}{d}; \bs(b) = (n, [d] \lc l); \bs' = \bs[b \mt (n, l)]$
  \[ \pbt{P, \bs} = (\nu \tilde{c_1} \tilde{c_2})( \pbt{Q} \pc \pbt{\bs\subs{\tilde{c}}{\nu \tilde{c}} \backslash b } \pc F_{n, [d_1, d_2] \lc L_l}(b_1, b_2) ) \]
  \[ \pbt{P, \bs} \xra{\op{b_1}{d_1, d_2}} (\nu \tilde{c_1} \tilde{c_2})( \pbt{Q} \pc \pbt{\bs\subs{\tilde{c}}{\nu \tilde{c}} \backslash b } \pc F_{n, L_l}(b_1, b_2) ) =  \pbt{P', \bs'} \]

  \tts{NewB*} $ P = (\nu b:n)P_1; P' = (\nu b)P_1; \alpha = \tau; \bs' = \bs[b \mt (n, l)] $ where $l = [\ ]$ \\
  \[ \begin{array}{l}
  \pbt{P, \bs} = (\nu \tilde{c_1} \tilde{c_2})\pbt{Q, \bs\subs{\tilde{c}}{\nu \tilde{c}}} \\
  \qquad \quad =(\nu \tilde{c_1} \tilde{c_2}) ( (\nu b_1 b_2)\tau.(\pbt{Q'} \pc F_{n, L_l}(b_1, b_2)) \pc \pbt{\bs\subs{\tilde{c}}{\nu \tilde{c}}} )
  \end{array} \]
  where $\{\tilde{c}\} = ln(\bs) \cup (\dom{\bs} \cap ln(P)), P \equiv_\bs (\nu \tilde{c})Q$ and $Q = (\nu b:n)Q'$.
  Since $b \not\in \dom{\bs}$, then $b\not\in n(\bs)$
  \[ \pbt{P, \bs} \xra{\tau} (\nu \tilde{c_1} \tilde{c_2})(\nu b_1 b_2)( \pbt{Q'} \pc F_{n, L_l}(b_1, b_2) \pc \pbt{\bs\subs{\tilde{c}}{\nu \tilde{c}}} ) = \pbt{P', \bs[b \mt (n, l)]} \]

  \tts{Sum} $ P = \sum_{i \in I}\pi_i.P_i; \pi_j.P_j, \bs \xra{\alpha} P', \bs'$. By induction,
  \[ \pbt{\pi_j.P_j, \bs} = \pbt{\pi_j.P_j} \pc \pbt{\bs} \xra{\at(\alpha)} \pbt{P', \bs'} \]
  \[ \pbt{P, \bs} = \sum_{i \in I}\pbt{\pi_i.P_i} \pc \pbt{\bs} \xra{\at(\alpha)} \pbt{P', \bs'} \]

  \tts{Par} $P = P_1 \pc P_2; P_1, \bs \xra{\alpha} P'_1, \bs'; P' = P'_1 \pc P_2 $. By induction
  \[ \pbt{P_1, \bs} = \pbt{P_1} \pc \pbt{\bs} \xra{\at(\alpha)} \pbt{P'_1, \bs'} = \pbt{P'_1} \pc \pbt{\bs'} \]
  Since new operators are guarded in $P_1 \pc P_2$, the transition does not involve any local names, hence $\bs'$ does not contain any local names either.
  \[ \pbt{P, \bs} = \pbt{P_1} \pc \pbt{P_2} \pc \pbt{\bs} \xra{\at(\alpha)} \pbt{P'_1} \pc \pbt{P_2} \pc \pbt{\bs'} = \pbt{P'_1 \pc P_2, \bs'} \]

  \tts{Com} $P = P_1 \pc P_2; P_1, \bs \xra{a(c)} P'_1, \bs; P_2, \bs \xra{\op{a}{c}} P'_2, \bs; P' = P'_1 \pc P'_2; \bs' = \bs $. By induction,
  \[ \pbt{P_1, \bs} = \pbt{P_1} \pc \pbt{\bs} \xra{\ip{a_1}{c_1, c_2}} \pbt{P'_1, \bs} = \pbt{P'_1} \pc \pbt{\bs} \]
  \[ \pbt{P_2, \bs} = \pbt{P_2} \pc \pbt{\bs} \xra{\op{a_2}{c_1, c_2}} \pbt{P'_2, \bs} = \pbt{P'_1} \pc \pbt{\bs} \]
  Hence
  \[ \pbt{P, \bs} = \pbt{P_1} \pc \pbt{P_2} \pc \pbt{\bs} \xra{\tau} \pbt{P'_1} \pc \pbt{P'_2} \pc \pbt{\bs} = \pbt{P'_1 \pc P'_2, \bs} \]

  \tts{New*} $P = (\nu c)P_1; P_1, \bs\subs{c}{\nu c} \xra{\alpha} P_2, \bs''; P' = (\nu c)P_2; \bs' = \bs''\subs{\nu c}{c} $. Suppose $c \not\in ln(\bs) \cup (\dom{\bs} \cap ln(P))$, the encoding is as follows
  \[ \pbt{P, \bs} = (\nu \tilde{d_1} \tilde{d_2})\pbt{Q, \bs\subs{\tilde{d}}{\nu \tilde{d}}} = (\nu \tilde{d_1} \tilde{d_2})( (\nu c_1 c_2)\pbt{Q'} \pc \pbt{\bs\subs{\tilde{d}}{\nu \tilde{d}}} ) \]
  where $\{\tilde{d}\} = ln(\bs)$, $P \equiv_\bs (\nu \tilde{d})Q$ and $Q = (\nu c)Q'$. Since $c \not\in gn(\bs)$, we can move the $(\nu c_1 c_2)$ to the outermost level.
  \[ \pbt{P, \bs} = (\nu c_1 c_2)(\nu \tilde{d_1} \tilde{d_2})( \pbt{Q'} \pc \pbt{\bs\subs{\tilde{d}}{\nu \tilde{d}}} ) = (\nu c_1, c_2)\pbt{P_1, \bs\subs{c}{\nu c}} \]
  By induction,
  \[ \pbt{P_1, \bs\{c/\nu c\}} \xra{\at(\alpha)} \pbt{P_2, \bs''} \]
  $c_1, c_2 \not\in n(\at(\alpha))$, since $c \not\in n(\alpha)$, then
  \[ \pbt{P, \bs} \xra{\at(\alpha)} (\nu c_1, c_2)\pbt{P_2, \bs''} = \pbt{(\nu c)P_2, \bs''\subs{\nu c}{c}} \]
  The last $=$ is because $c \not\in gn(\bs''\subs{\nu c}{c})$ and $c \not\in ln(\bs'')$.

  \tts{Open*} $ P = (\nu c)P''; \bs' = \bs\subs{c}{\nu c}; P'', \bs' \xra{\op{d}{c}} P', \bs'; \alpha = \op{d}{\nu c}$ \\
  As in \tts{New*},
  \[ \pbt{P, \bs} = (\nu c_1, c_2)\pbt{P'', \bs'} \]
  By induction,
  \[ \pbt{P'', \bs'} \xra{\at(\op{d}{c})} \pbt{P', \bs'} \]
  \[ \pbt{P, \bs} \xra{\at(\op{d}{\nu c})} \pbt{P', \bs'} \]

  \tts{Stru} The result follows from Lemma~\ref{lem:sctrans} \qed
\end{proof}

Proof of Lemma~\ref{lem:gosim2}

\begin{proof}
  We prove by induction on the size of $P$. Consider the structure of $P$:

  For input prefix. $P = \ip{c}{x}.P'$ and $ln(\bs) = \emptyset$.
  \[ \pbt{P, \bs} = \ip{c_1}{x_1, x_2}.\pbt{P'} \pc \pbt{\bs} \xra{\at(\alpha)} R \]
  If $c \not\in \dom{\bs}$,
  \[ \pbt{P, \bs} \xra{\ip{c_1}{d_1, d_2}} \pbt{P'}\subs{d_1, d_2}{x_1, x_2} \pc \pbt{\bs} = \pbt{P'\subs{d}{x}, \bs} \]
  If $c \in \dom{\bs}$, the encoding may also perform this action.
  However, only the buffer process $F_{n, L}(c_1, c_2)$ is able to perform the complementary $\xra{\op{c_1}{d_1, d_2}}$ action,
  hence if $c$ is a buffered names, we only need to consider the following transition:
  \[ \pbt{P, \bs} \xra{\tau} \pbt{P'}\subs{d_1, d_2}{x_1, x_2} \pc F_{n, L_l}(c_1, c_2) \pc \pbt{\bs\backslash c}) = \pbt{P'\subs{d}{x}, \bs[c \mt (n, l)]} \]
  where $\bs(c) = (n, [d] \lc l)$.

  For output prefix. $P = \op{c}{d}.P'$ and $ln(\bs) = \emptyset$.
  \[ \pbt{P, \bs} = \op{c_2}{d_1, d_2}.\pbt{P'} \pc \pbt{\bs} \xra{\at(\alpha)} R \]
  If $c \not\in \dom{\bs}$,
  \[ \pbt{P, \bs} \xra{\op{c_2}{d_1, d_2}} \pbt{P'} \pc \pbt{\bs} = \pbt{P', \bs} \]
  If $c \in \dom{\bs}$ and $\bs(c) = (n, l)$ where $\len{l} < n$
  \[ \pbt{P, \bs} \xra{\tau}  \pbt{P'} \pc F_{n, L_l \lc [(d_1, d_2)]}(c_1, c_2) \pc \pbt{\bs\backslash c} = \pbt{P', \bs[c \mt (n, l \lc [d])]} \]

  For summation. $P = \sum_{i \in I}P_i$ and $ln(\bs) = \emptyset$.
  \[ \pbt{P, \bs} = \sum_{i \in I}\pbt{P_i} \pc \pbt{\bs} \xra{\at(\alpha)} R \]
  This $\at(\alpha)$ may be an action of $\sum_{i \in I}\pbt{P_i}$ alone, or communication between $\sum_{i \in I}\pbt{P_i}$ and $\pbt{\bs}$. In any case, suppose $\pbt{P_j} \xra {\alpha'} R'$, then
  \[ \pbt{P_j, \bs} = \pbt{P_j} \pc \pbt{\bs} \xra{\at(\alpha)} R \]
  By induction, $P_j, \bs \xra{\alpha} P'_j, \bs'$ and $R = \pbt{P'_j, \bs'}$. By \tts{Sum}
  \[ \sum_{i \in I}P_i, \bs \xra{\alpha} P'_j, \bs' \]

  For parallel composition. $P = P_1 \pc P_2$ and $ln(\bs) = \emptyset$
  \[ \pbt{P, \bs} = \pbt{P_1} \pc \pbt{P_2} \pc \pbt{\bs} \xra{\at(\alpha)} R \]
  This $\at(\alpha)$ may be induced by $\pbt{P_1}$(or $\pbt{P_2}$) alone, or communication between $\pbt{P_1}$ and $\pbt{P_2}$. For the former, suppose
  \[ \pbt{P_1, \bs} = \pbt{P_1} \pc \pbt{\bs} \xra{\at(\alpha)} R' \]
  By induction
  \[ P_1, \bs \xra{\alpha} P'_1, \bs' \text{  and  } R' = \pbt{P'_1, \bs'}\]
  Consider the following cases regarding $\bs'$: if $c \in ln(\bs')$, this means $P_1$ sends a local name $c$ to  $\bs$. Then $P_1 \equiv_{\bs} (\nu c)P''_1$, $P'_1 \equiv_\bs (\nu c)P'''_1$ and
  \[ \pbt{P_1, \bs} \xra{\at(\alpha)} (\nu c_1, c_2)(\pbt{P_1'''} \pc \pbt{\bs'\subs{c}{\nu c}}) \]
  Suppose $c$ does not occur in $P_2$ (if it does, rename $c$ to a name not occur in $P_2$), we have
  \[ \pbt{P, \bs} \xra{\at(\alpha)} (\nu c_1, c_2)(\pbt{P_1'''} \pc \pbt{P_2} \pc \pbt{\bs'\subs{c}{\nu c}}) = \pbt{P'', \bs'} \]
  \[ P_1 \pc P_2 \equiv_{\bs} (\nu c)(P_1'' \pc P_2) \quad (\nu c)(P_1'' \pc P_2), \bs \xra{\alpha} P'', \bs' \]
  where $P'' = (\nu c)( P_1''' \pc P_2)$. If $ln(\bs') = \emptyset$, then
  \[ \pbt{P_1, \bs} \xra{\at(\alpha)} \pbt{P_1'} \pc \pbt{\bs'} \]
  \[ \pbt{P, \bs} \xra{\at(\alpha)} \pbt{P_1'} \pc \pbt{P_2} \pc \pbt{\bs'} = \pbt{P'_1 \pc P_2, \bs'} \]
  \[ P_1 \pc P_2, \bs \xra{\alpha} P'_1 \pc P_2, \bs' \]
  If $\at(\alpha)$ is a communication action between $\pbt{P_1}$ and $\pbt{P_2}$. Suppose
  \[ \pbt{P_1, \bs} \xra{a_1(c_1, c_2)} R_1 \quad \pbt{P_2, \bs} \xra{\op{a_2}{c_1, c_2}} R_2 \]
  By induction
  \[ P_1, \bs \xra{a(c)} P'_1, \bs \quad P_2, \bs \xra{\op{a}{c}} P'_2, \bs \text{ and } R_1 = \pbt{P'_1, \bs} \quad R_2 = \pbt{P'_2, \bs} \]
  Therefore
  \[ P_1 \pc P_2, \bs \xra{\tau} P'_1 \pc P'_2, \bs \]
  \[ \pbt{P_1 \pc P_2, \bs} \xra{\tau} \pbt{P'_1} \pc \pbt{P'_2} \pc \pbt{\bs} = \pbt{P'_1 \pc P'_2, \bs} \]

  For new process. $P = (\nu a)P'$, Suppose $a \not\in gn(\bs)$
  \[ \pbt{P, \bs} \equiv (\nu a_1, a_2)(\pbt{P', \bs\subs{a}{\nu a}}) \xra{\at(\alpha)} R \]
  If $a \not\in n(\alpha)$, we have
  \[ \pbt{P', \bs\subs{a}{\nu a}} \xra{\at(\alpha)} R' \]
  By induction, $ P', \bs\subs{a}{\nu a} \xra{\alpha} P'', \bs' $ and $R' = \pbt{P'', \bs'}$. Since $R = (\nu a_1, a_2)R' = (\nu a_1, a_2)\pbt{P'', \bs'}$. Suppose $a \not\in ln(\bs')$, by rule \tts{New*}
  \[ (\nu a)P', \bs \xra{\alpha} (\nu a)P'', \bs'\subs{\nu a}{a} \]
  \[ R = \pbt{(\nu a)P'', \bs'\subs{\nu a}{a}} \]
  Suppose $a \in ln(\bs')$, this means $P'$ outputs a new name $a$ (not the outermost $\nu a$ of $P$) to the buffer, then $a \not\in ln(\bs)$ since $a \not\in gn(\bs\subs{a}{\nu a})$, also $a$ is not a free name of $P'$. Choose a fresh name $a'$ such that $\nt(a') = (a'_1, a'_2)$ and $a'_1, a'_2$ are fresh names in $\pbt{P, \bs}$, then $R \equiv (\nu a'_1, a'_2)\pbt{P'', \bs'}$ \\
  \[ (\nu a)P' \equiv_{\bs} (\nu a')P' \]
  \[ (\nu a')P', \bs \xra{\alpha} (\nu a')P'', \bs' \]
  \[ R = \pbt{(\nu a')P'', \bs'} \]
  If $a \in n(\alpha)$ i.e. $\alpha = \op{a'}{\nu a}$, we have
  \[ \pbt{P', \bs\subs{a}{\nu a}} \xra{\at(\op{a'}{a})} R \]
  By induction, $R = \pbt{P'', \bs\subs{a}{\nu a}}$ and $ P', \bs\subs{a}{\nu a} \xra{\op{a'}{a}} P'', \bs\subs{a}{\nu a} $. By rule \tts{Open*}
  \[ (\nu a)P', \bs \xra{\alpha} P'', \bs\subs{a}{\nu a} \]
  Suppose $a \in gn(\bs)$ ($a \not\in ln(\bs)$), the above equations are also valid after appropriate $\alpha$-conversiton of $(\nu a_1, a_2)$, and this $\alpha$-conversion corresponds to renaming of $a$ in $P$.
  \[ (\nu a)P' \equiv_{\bs} (\nu a')P'\subs{a'}{a} \]
  where $a'$ is a fresh name.

  For extended new process. $P = (\nu b:n)P'$.
  \[ \pbt{P, \bs} =(\nu \tilde{d_1} \tilde{d_2}) ( (\nu b_1 b_2)\tau.(\pbt{Q'} \pc F_{n, L_l}(b_1, b_2)) \pc \pbt{\bs\subs{\tilde{d}}{\nu \tilde{d}}} ) \xra{\at(\alpha)} \]
  where $l = [\ ]$. The process the following transition
  \[ \pbt{P, \bs} \xra{\tau} (\nu \tilde{d_1} \tilde{d_2}) ( (\nu b_1, b_2)(\pbt{Q'} \pc F_{n, L_l}(b_1, b_2)) \pc \pbt{\bs\subs{\tilde{d}}{\nu \tilde{d}}} ) \]
  Suppose $b \not\in \texttt{dom}(\bs)$, then $ b \not\in n(\bs)$,
  \[ P, \bs \xra{\tau} P', \bs[b \mt (n, l)] \]
  \[ \pbt{P, \bs} \xra{\tau} (\nu \tilde{d_1} \tilde{d_2})(\nu b_1, b_2)( \pbt{Q'} \pc F_{n, L_l}(b_1, b_2) \pc \pbt{\bs\subs{\tilde{d}}{\nu \tilde{d}}} ) = \pbt{P', \bs[b \mt (n, l)]} \]
  Suppose $b \in \dom{\bs}$, the equations are also valid after appropriate $\alpha$-conversion of $(\nu b_1 b_2)$, and this $\alpha$-conversion corresponds to renaming of $b$ in $P$.

  For replication. $P = !P' \equiv P' \pc !P'$ and $ln(\bs) = \emptyset$.
  \[ \pbt{!P', \bs} = !\pbt{P'} \pc \pbt{\bs}  \equiv_{\bs}   \pbt{P'} \pc !\pbt{P'} \pc \pbt{\bs} = \pbt{P' \pc !P, B } \]

  For any process $P$, if $c \in \dom{\bs}$ and $c \not\in ln(P)$, then
  \[ \pbt{P, \bs} = (\nu \tilde{d_1} \tilde{d_2})(\pbt{Q} \pc \pbt{\bs\subs{\tilde{d}}{\nu \tilde{d}}\backslash c} \pc F_{n, L_l}(c_1, c_2)) \xra{\at(\alpha)}  \]
  Suppose $\bs(c) = (n, l)$ and $\len{l} < n$, then
  \[ \pbt{P, \bs} \xra{\ip{c_2}{d_1, d_2}} (\nu \tilde{d_1} \tilde{d_2})(\pbt{Q} \pc \pbt{\bs\subs{\tilde{d}}{\nu \tilde{d}}\backslash c} \pc F_{n, L_l \lc [(d_1, d_2)]}(c_1, c_2))  \]
  \[ P, \bs \xra{\ip{c}{d}} P, \bs[c \mt (n, l \lc [d])] \]
  Suppose $\bs(c) = (n, [d] \lc l')$, then
  \[ \pbt{P, \bs} \xra{\op{c_1}{d_1, d_2}} (\nu \tilde{d_1} \tilde{d_2})(\pbt{Q} \pc \pbt{\bs\subs{\tilde{d}}{\nu \tilde{d}}\backslash c} \pc F_{n, L_{l'}}(c_1, c_2))  \]
  \[ P, \bs \xra{\op{c}{d}} P, \bs[c \mt (n, l')] \]

  This completes the proof. \qed
\end{proof}

\end{document}